\documentclass[aps,prl,reprint,superscriptaddress]{revtex4-1}

\usepackage[utf8]{inputenc}
\usepackage{graphicx}
\usepackage{amsmath}
\usepackage{bm}
\usepackage{dcolumn}
\usepackage{mhchem}
\usepackage[separate-uncertainty=true]{siunitx}
\usepackage{xcolor}
\usepackage{hyperref}
\usepackage{cleveref}
\usepackage[english]{babel}
\usepackage[autostyle, english=american]{csquotes}
\usepackage{newunicodechar}
\newunicodechar{ }{\,}
\begin{document}

\onecolumngrid

\begin{center}
{\Large\bfseries 
Quantifying the Full Damage Profile of Focused Ion Beams via 4D-STEM Precession Electron Diffraction and PSNR Metrics}
\vspace{1em}

M.G. Masteghin\textsuperscript{1,2,*},
Z.P. Aslam\textsuperscript{3},
A.P. Brown\textsuperscript{3},
M.J. Whiting\textsuperscript{4},
S.K. Clowes\textsuperscript{1},
R.P. Webb\textsuperscript{5},
D.C. Cox\textsuperscript{1,5,6}

\vspace{1em}

\textsuperscript{1} Advanced Technology Institute, University of Surrey, Guildford, GU2 7XH, UK \\
\textsuperscript{2} DTU Nanolab, Technical University of Denmark, Fysikvej, Kongens Lyngby 2800, Denmark \\
\textsuperscript{3} School of Chemical and Process Engineering, University of Leeds, Leeds, LS2 9JT, UK \\
\textsuperscript{4} School of Mechanical Engineering Sciences, University of Surrey, Guildford, GU2 7XH, UK \\
\textsuperscript{5} Ion Beam Centre, University of Surrey, Guildford GU2 7XH, UK \\
\textsuperscript{6} National Physical Laboratory, Teddington, TW11 0LW, United Kingdom \\
\textsuperscript{*} Corresponding author: \href{mailto:matmas@dtu.dk}{matmas@dtu.dk}
\end{center}

\date{\today}

\vspace{1em}

\begin{abstract}
\noindent
Focused ion beams (FIBs) are widely used in nanofabrication for applications such as circuit repair, ultra-thin lamella preparation, strain engineering, and quantum device prototyping. Although the lateral spread of the ion beam is often overlooked, it becomes critical in precision tasks such as impurity placement in host substrates, where accurate knowledge of the ion-matter interaction profile is essential. Existing techniques typically characterise only the beam core, where most ions land, thus underestimating the full extent of the point spread function (PSF). In this work, we use four-dimensional scanning transmission electron microscopy (4D-STEM) to resolve the ion beam tail at defect densities equivalent to $<$0.1 ions nm$^{-2}$. Convergent beam electron diffraction (CBED) patterns were collected in calibration regions with known ion fluence and compared to patterns acquired around static dwell spots exposed to a 30~keV Ga$^{+}$ beam for 1–10~s. Cross-correlation using peak signal-to-noise ratio (PSNR) revealed that 4D-STEM datasets are ultra-sensitive for defect quantification and more robust against scanning artefacts than conventional dark-field imaging. This approach is complementary to image resolution methods enabling a comprehensive profiling of ion-induced damage even at low-dose regimes, offering a more accurate representation of FIB performance and supporting application tailoring beyond the conventional resolution metrics focused solely on the beam core.
\end{abstract}

\maketitle
\begingroup
\renewcommand\thefootnote{\fnsymbol{footnote}}
\footnotetext[1]{Corresponding author: \href{mailto:matmas@dtu.dk}{matmas@dtu.dk}}
\endgroup

\twocolumngrid

\section{Introduction}
Focused ion beam (FIB) systems are indispensable tools in nanofabrication, offering high spatial resolution for ion implantation \cite{McCallum2012,Adshead2023,Masteghin2024}, milling \cite{Mayer2007}, and other material modifications such as strain engineering \cite{Masteghin2021} or composite deposition \cite{Masteghin2018}. In advanced applications such as quantum devices fabrication, where atomic-scale precision is paramount \cite{Kane1998}, understanding the full spatial extent of ion-induced damage becomes crucial. This is particularly important in the low-dose regime, where even very limited ion exposure can degrade the coherence properties of quantum bits (qubits) \cite{Masteghin2024b,Holmes2024} or alter the optical properties of quantum dots \cite{Lee2011}. Accurate characterisation of ion beam damage profiles not only could reveal intrinsic limitations of FIB techniques but also provide a foundation for refining microscope optics, developing mitigation strategies, and tailoring tool performance to meet the stringent requirements of emerging quantum technologies \cite{Masteghin2024,Schofield2025}.

Drezner et al. \cite{Drezner2017} discussed several techniques to measure ion beam profile and their limitations, such as scanning across heterostructures, knife-edge methods, and resist-based imaging. In the heterostructure approach, the beam is scanned across a material interface (e.g. Si/SiGe), and the resulting contrast (or grey levels) is used to infer the beam shape. This approach is discussed below. While this method provides good lateral resolution, it requires complex sample preparation and is sensitive to interface quality. Knife-edge techniques involve scanning the beam across a sharp boundary (e.g. a metal edge or a patterned mask) and analysing the resulting signal gradient. These methods are straightforward, but often suffer from convolution effects and limited sensitivity to low-dose tails. Resist-based imaging, where ion exposure modifies a resist layer, offers high contrast but lacks quantitative accuracy and is limited by resist sensitivity thresholds. Hence, most techniques are only accurate for determination of the full width at half maximum of the ion beam “core”, or what we will refer to here as “imaging resolution”.

To overcome these limitations, we present a novel application of four-dimensional scanning transmission electron microscopy (4D-STEM). We propose the use of 4D-STEM to map ion beam damage with high sensitivity and spatial resolution in the low-dose regime. In 4D-STEM, a full convergent beam electron diffraction (CBED) pattern is recorded at each scan position, enabling detailed analysis of the local crystal structure. By comparing CBED patterns from pristine and Ga$^+$-implanted regions (30 keV, doses from 0.09 to 5 ions nm$^{-2}$), we extract peak signal-to-noise ratio (PSNR) values as a quantitative metric of structural degradation. These values are then mapped across a 30 nm thick silicon crystal irradiated at a single point for varying durations, and the pixels position for the surrounding area are converted to local ion densities using a calibration curve derived from the known-dose regions. This method is uniquely suited to probe the beam tail: while it is insensitive to the highly damaged beam core, where diffraction contrast is lost, it is exceptionally sensitive to subtle structural changes in the low-dose periphery. This makes it ideal for applications where understanding the spatial extent of minimal damage is crucial, such as in deterministic ion implantation for quantum technologies.

Recent advances have positioned 4D-STEM as a powerful and increasingly popular technique for assessing crystal quality. It has been used to map strain \cite{Mahr2021}, identify defects \cite{Fang2019}, and reconstruct phase information with atomic resolution \cite{Chen2021}. Its ability to capture rich diffraction data at each scan point enables new insights into local structure and disorder, making it a valuable tool in both fundamental research and applied materials science.  Our work extends the application portfolio for 4D-STEM, including the use of PSNR-based analysis for quantitative ion beam point spread function profiling. This approach complements existing techniques and provides a robust, high-resolution method for mapping ion beam tails, an area where traditional methods often fall short.

\section*{Materials and Methods}
The 30~nm thick single-crystal Si membranes were commercially manufactured by Silson Ltd. The window edges and diagonals are oriented along the $<110>$ and $<100>$ directions, respectively, with (001) surface plane. Details on how Si membranes can be produced can be found in \cite{Palik1988,vu1996,zubel2000}.

Singly charged gallium ions ($Ga^{+}$) with an energy of 30 keV were implanted using a dual-beam focused ion beam microscope and scanning electron microscope (FIB-SEM, FEI Nova Nanolab). The gallium ion current was measured before and after implantation using a Faraday cup and determined to be 7.5~pA on both readings and, therefore, was assumed constant throughout the implantation. The conventional resolution test was performed by imaging a gold on carbon sample, capturing an image of $1024 \times 1024$~pixels$^2$ with a dwell time of \SI{30} {\micro\second}. For dark-field scanning transmission electron microscopy (STEM) imaging, a line of milled spots in a 30~nm thick Si membrane was created by scanning the ion beam with a pitch of 5000~nm and dwell times exponentially increasing from 0.08~ms to 500~ms. For the 4D-STEM measurements, the patterns consisted of two main components, the calibration squares and the milled spots. The 28 calibration squares had dimensions equal to $2500 \times 2500$~nm$^2$ with a centre-to-centre spacing of 3000~nm. The ion dose in each square was determined by the dwell time for each scanned pixel (with a pitch size of 10.7~nm), varying exponentially (dose~$=1.182\times1,176^j$, being $j$ the index from 1 to 26) from \SI{1.39}{\micro\coulomb\per\centi\meter\squared} (repeated for three squares) to \SI{80}{\micro\coulomb\per\centi\meter\squared}.
A $10^{-6} \times (\text{cm}^{-2} / ((10^7)^2\, \text{nm}^2)) \times (1 / (1.6 \times 10^{-19}\, \text{C}))$ conversion results in doses ranging from 0.09~ions~nm$^{-2}$ to 5.00~ions nm$^{-2}$.
For the milled spots, the beam dwelled a fixed pixel for 1~s, 5~s, and 10~s. Additional implantation was carried out to generate alignment marks, L-shaped and triangular, with doses of \SI{80}{\micro\coulomb\per\centi\meter\squared} and \SI{1600}{\micro\coulomb\per\centi\meter\squared}, respectively, as well as to flatten the membrane and mitigate out-of-plane buckling as suggested in \cite{Masteghin2021, Masteghin2024b}.
The sample was not annealed, i.e., no attempt was made to activate the Ga atoms and any damage due to the implantation was not repaired.

Annular dark-field (ADF) images were acquired using a Thermo Scientific Talos F200i operated in STEM mode at \SI{200}{\kilo\electronvolt}, at “spot-size” 5, with a selected intermediate condenser aperture size of \SI{70}{\micro\meter}, resulting in a convergence semi-angle of around 10.5~mrad. The resulting screen current was $\approx$\SI{200}{\pico\ampere} and the $2048\times512$ pixels$^2$ images were obtained using a \SI{10}{\micro\second} dwell time. The holes and damage diameters were measured using the internodes thresholding method in ImageJ. The ADF-STEM images related to the 4D-STEM experiments were acquired in the Tescan TENSOR.

Diffraction datasets were acquired using a Tescan TENSOR 4D-STEM microscope operated at an accelerating voltage of 100~keV. During scanning, the electron beam was dynamically precessed, and diffraction patterns (DPs) were recorded using a DECTRIS Quadro hybrid-pixel direct electron detector. Two experimental configurations were employed to optimize data quality and spatial resolution. In the first configuration, the beam was precessed at 0.8~mrad, with a convergence semi-angle of 12~mrad and a beam current of 100~pA. Under these conditions, diffraction patterns were acquired with a dwell time of 1~ms per scan point. In the second configuration, the beam was precessed at 0.7~mrad, with a convergence semi-angle of 8~mrad and a beam current of 210~pA. A longer dwell time of 10~ms was used to enhance signal acquisition.

Kikuchi patterns were acquired using a JEOL JSM-7100F field emission scanning electron microscope (SEM) equipped with a Thermo Fisher Lumis EBSD detector. The region of interest was scanned with a 20~keV electron beam at a current of 0.95~nA, using a dwell time of 50~ms and a step size of ~150~nm.

\section*{Results and Discussion}

\begin{figure}[t]
    \centering
    \includegraphics[width=0.8\linewidth]{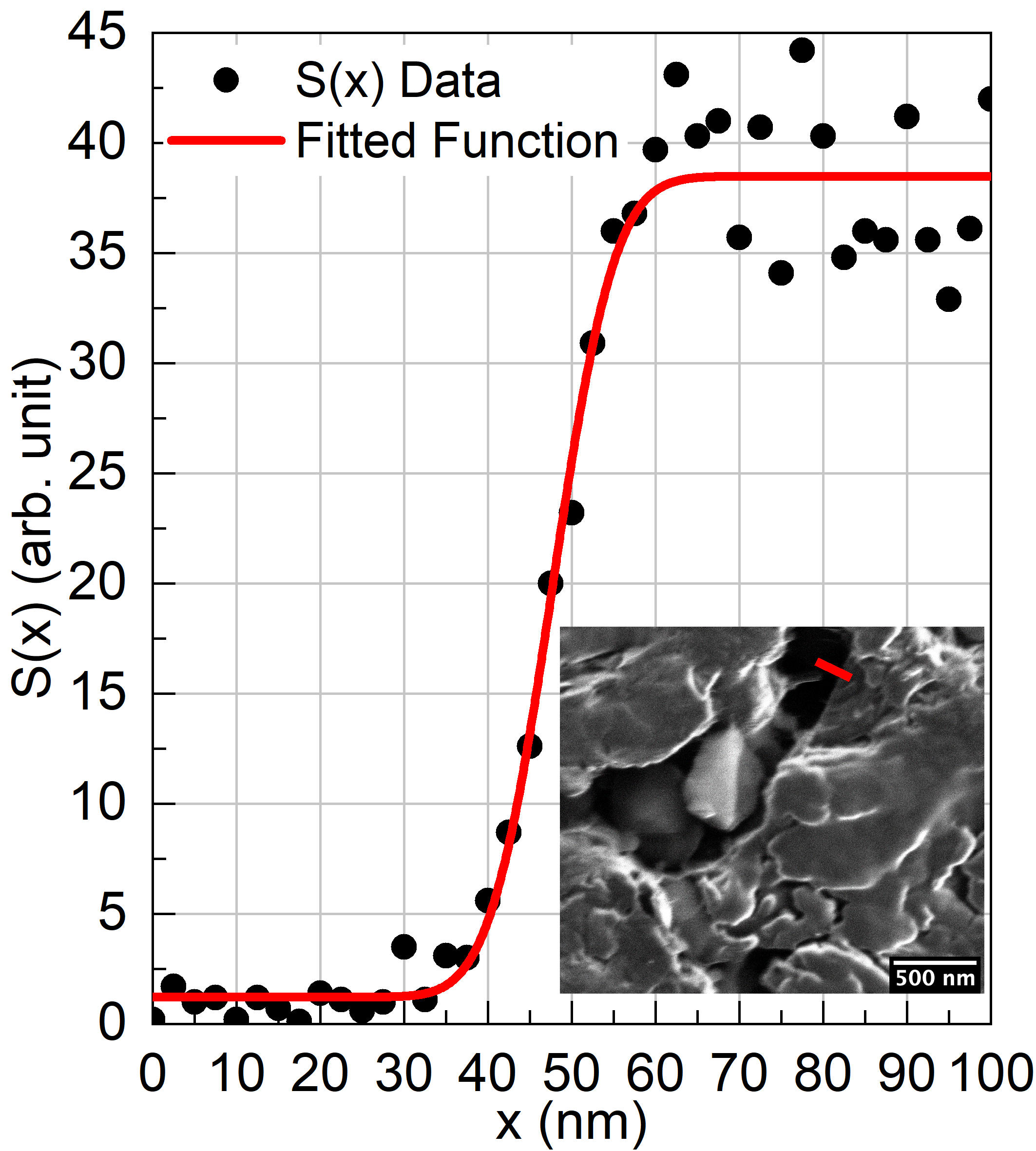}
    \caption{Conventional imaging resolution measurement using linear greyscale profile across a heterojunction. Inset: A $1024 \times 943$ pixel$^2$ secondary electrons micrograph obtained by scanning a 30~keV Ga$^+$ beam such as around 4200 electrons per pixel is used to produce the image. The red line on the top-middle part of the inset figure represents the region where a greyscale profile $S(x)$ was collected. The data was fitted using \cref{Eq.Sx} to give imaging resolution $2\sigma=11.9$~nm. }
    \label{fig:ImagingResolution}
\end{figure}

FIB manufacturers started adopting knife-edge measurement as a standard to assess their system resolution. This method consists of moving either a solid material under the beam (e.g. nanomanipulator) or scanning the beam across a heterogeneous interface \cite{Trgrdh2015}. This technique assumes a perfectly circular beam (i.e., no stigmatism) with a Gaussian intensity distribution \cite{Arimoto1983, Svelto2023}.  So in the case of a solid knife edge in position $x_\text{e}$ and a beam centred on the origin, the measured beam current is given by \cite{Arnaud1971, Suzaki1975}
\begin{multline}
    I(x_\text{e})=\int_{x_\text{e}}^{+\infty}\int_{-\infty}^{+\infty}I_{(x,y)}dxdy= \\
    \int_{x_\text{e}}^{+\infty}\int_{-\infty}^{+\infty}\frac{I_0}{2\pi\sigma^2}e^{-\frac{x^2+y^2}{2\sigma^2}}dxdy=\frac{I_0}{2}\text{erfc}\left[\frac{x_\text{e}}{\sqrt{2}\sigma}\right],
\end{multline}
where $I_0$ is the total beam current, $\sigma$ is the standard deviation of the beam width, and erfc is the complementary error  function. For the case of scanning the beam across a heterogeneous interface, where the secondary electron emission is $K$ multiplied by the secondary electron yield for each side of the interface ($\gamma$ and $\delta$), the measured greyscale $S$ plot as a function of beam position is given by
\begin{equation}
    S(x)=K\left[\frac{\gamma+\delta}{2}+\frac{\gamma-\delta}{2}\text{erf}\left(\frac{x-x_i}{\sqrt{2}\sigma}\right)\right]
    \label{Eq.Sx}
\end{equation}
where $x_i$ is the position of the interface and erf is the standard error function. For both methods, the quoted imaging resolution is $2\sigma$, which is often referred to as the ``spot size'' or ``probe size''. An example of the latter method is shown in \cref{fig:ImagingResolution}, in which the inset shows a secondary electron image obtained after scanning a 30~keV, 7.5~pA Ga$^+$ beam across a heterogeneous gold/carbon interface, and the greyscale profile $S(x)$ is extracted across the region marked by the red line in the image. In this example, from fitting $S(x)$ to \cref{Eq.Sx}, the imaging resolution of $2\sigma=11.9$~nm is obtained.

A comprehensive summary of the challenges associated with using these methods is provided by Drezner et al.\cite{Drezner2017}. When a wire is imaged whilst positioned on the top of a Faraday cup, the wire roughness is typically larger than the beam radius. Conversely, when a smooth nanomanipulator is moved between the pole-piece and the Faraday cup, the uncertainty of the piezo movement is often comparable to the beam point spread function. Therefore, in neither operational mode is the condition of a beam profile smaller than the transition regions met. In the alternative approach of scanning the beam across a heterogenous interface, the method risks interface smothering due to material sputtering, which must be mitigated by a fast dwell time. Nonetheless, when dwelling a 7.5~pA beam for approximately 100~ns, less than 3 electrons per pixel are generated \cite{Baragiola1979, Ohya2002}, resulting in a low signal-to-noise ratio and likely obscuring the beam tails.

\begin{figure*}[t]
    \centering
    \includegraphics[width=0.8\linewidth]{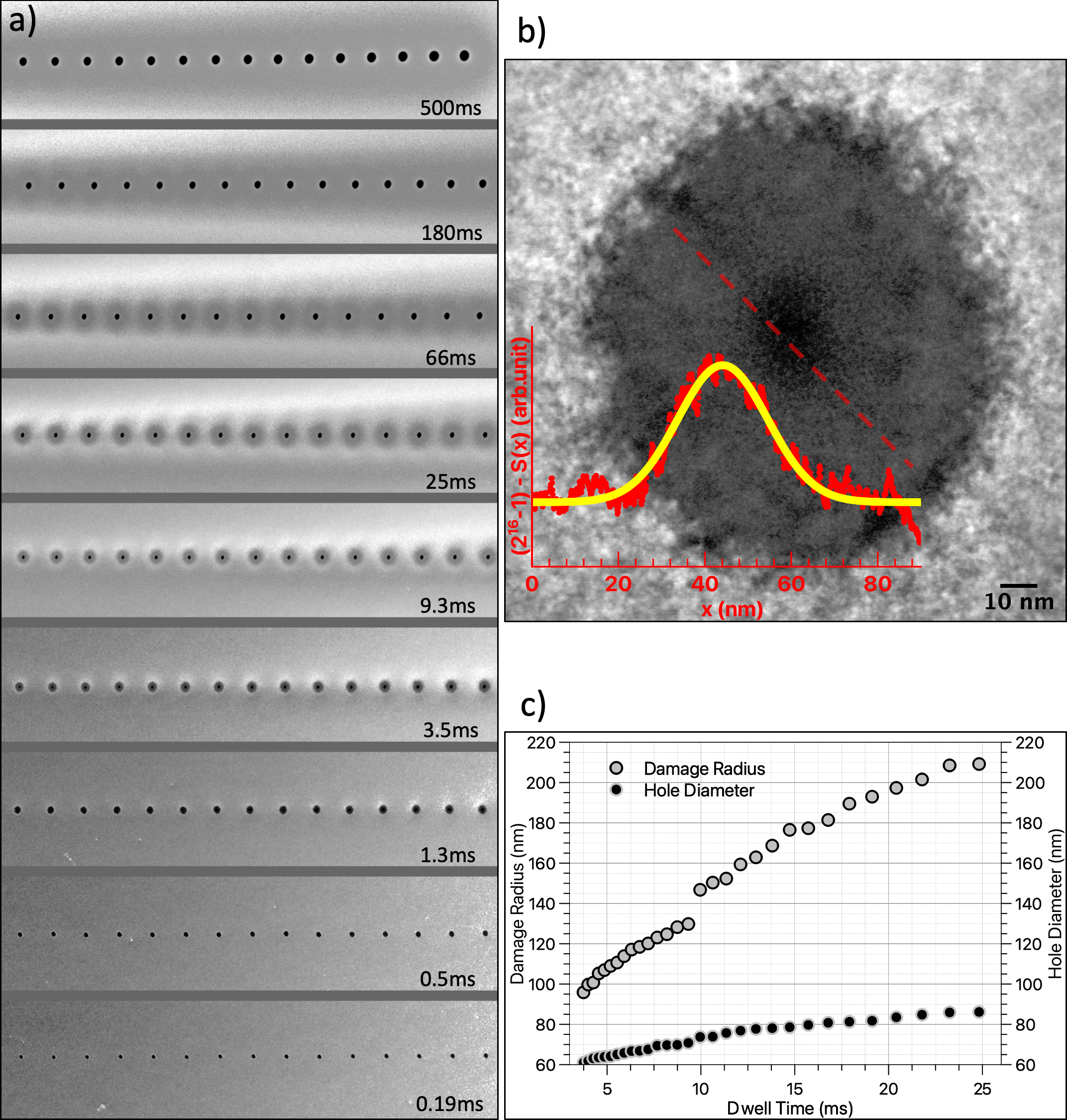}
    \caption{Imaging resolution versus damage profile study using annular dark-field scanning transmission electron microscopy (ADF-STEM). a) ADF-STEM images around spots dwelled with a 7.5~pA 30~keV Ga$^+$ beam. The times on the bottom-right of each image is the dwell time for the far-right spot, with exponential increments between them. Scan field of view corresponds to \SI{8}{\micro\meter}. b) High-magnification high-angle annular dark-field (HAADF) STEM image of the initially drilled hole. The image shows a central darker region, corresponding to the hole or very thin areas, surrounded by a dark-grey annular region, indicative of damaged (likely amorphous) areas. The untouched crystal appears in light grey. The red dashed line marks the region where the greyscale profile was obtained, which is plotted in the red inset. The yellow curve represents a Gaussian fit with a full width at half maximum equivalent to the imaging resolution obtained in \cref{fig:ImagingResolution}. c) Evolution of hole size and the corresponding surrounding damage, demonstrating a progressively higher damage profile relative to the hole diameter.}
    \label{fig:ADF-Holes}
\end{figure*}

This can be easily demonstrated with a simple STEM experiment. A 30~keV, 7.5~pA Ga$^+$ beam was dwelled in spot mode for times ranging exponentially from 0.08 to 500~ms. The 30~nm-thick silicon single-crystal membrane was oriented along the $\left<001\right>$ zone axis and imaged using the annular dark-field (ADF) detector (\cref{fig:ADF-Holes}a). When the first hole was observed ($\approx$10~ms), it was imaged at a higher magnification, and a greyscale profile across the darker central region (taken across the red dashed line) was obtained (\cref{fig:ADF-Holes}b). A Gaussian fit of $S(x)$ resulted in a full width at half-maximum (FWHM) of around 15~nm. It is evident that the imaging resolution corresponds to the region where most ions arrive, leading to increased material sputtering. Substantial damage occurs in the area surrounding the dwelled pixel, as highlighted in \cref{fig:ADF-Holes}c, which presents plots of damage radii and hole diameter as a function of dwell times. The majority of ions is concentrated near the centre of the spot, leading to rapid material removal and hole formation. However, ions arrivals increasingly contribute to damage in the surrounding regions due to the broader tails of the beam profile and scattering effects such as transversal collision cascades. As a result, the damage continues to accumulate laterally as the rate of hole diameter growth slows down, leading to a steeper increase in overall damage compared to hole size. Hence, although the focused ion beam has a narrow core responsible for milling, its spatial profile includes extended reach that contributes to ion-matter interactions. These interactions are responsible for the creation of point defects \cite{Lesik2013, Schofield2025}, amorphisation \cite{Lugstein2003, Rubanov2004}, and strain generation\cite{Masteghin2021}. Therefore, the ion beam community would benefit from a highly sensitive metrology framework capable of characterising beam tail effects in nanoscale FIB patterning.

Based on the data in \cref{fig:ADF-Holes}, STEM imaging alone should be sufficiently sensitive to estimate the extent of the ion beam tail. We, therefore, performed an ADF-STEM experiment to calibrate greyscale contrast as a function of ion dose (\cref{fig:damage_calib}). However, due to optical off-axis artefacts, a gradual offset in greyscale levels was observed across the scan, see supplementary materials. To overcome these limitations, we propose a 4D-STEM approach that would provide a comprehensive dataset at each scan position, enabling advanced post-processing (e.g. centre-of-mass correction) to mitigate such artefacts.

\begin{figure}[t]
    \centering
    \includegraphics[width=\linewidth]{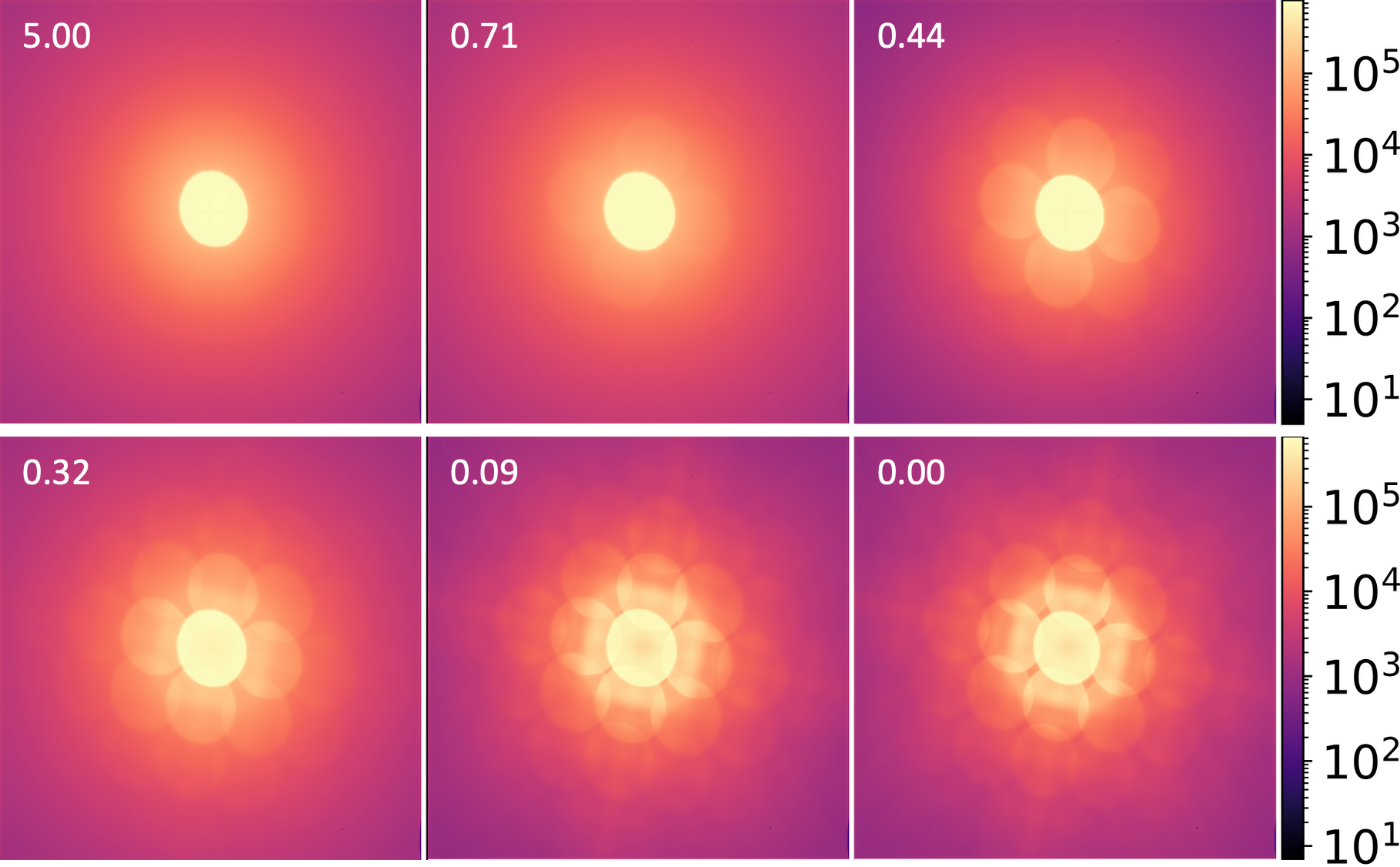}
    \caption{10,000 integrated CBED patterns obtained by scanning the 0.8~mrad precessed 100~keV, 12~mrad convergence, 100~pA probe current with a 1~ms dwell time per pixel. The CBED patterns have dimensions of $512\times512$ pixels$^2$ and are plotted using a logarithmic intensity scale. The numbers at the top-left of each CBED pattern indicate the dose used in the corresponding calibration square, with 0.00 representing the pristine region.}
    \label{fig:CBED_series}
\end{figure}

\begin{figure}[b]
    \centering
    \includegraphics[width=\linewidth]{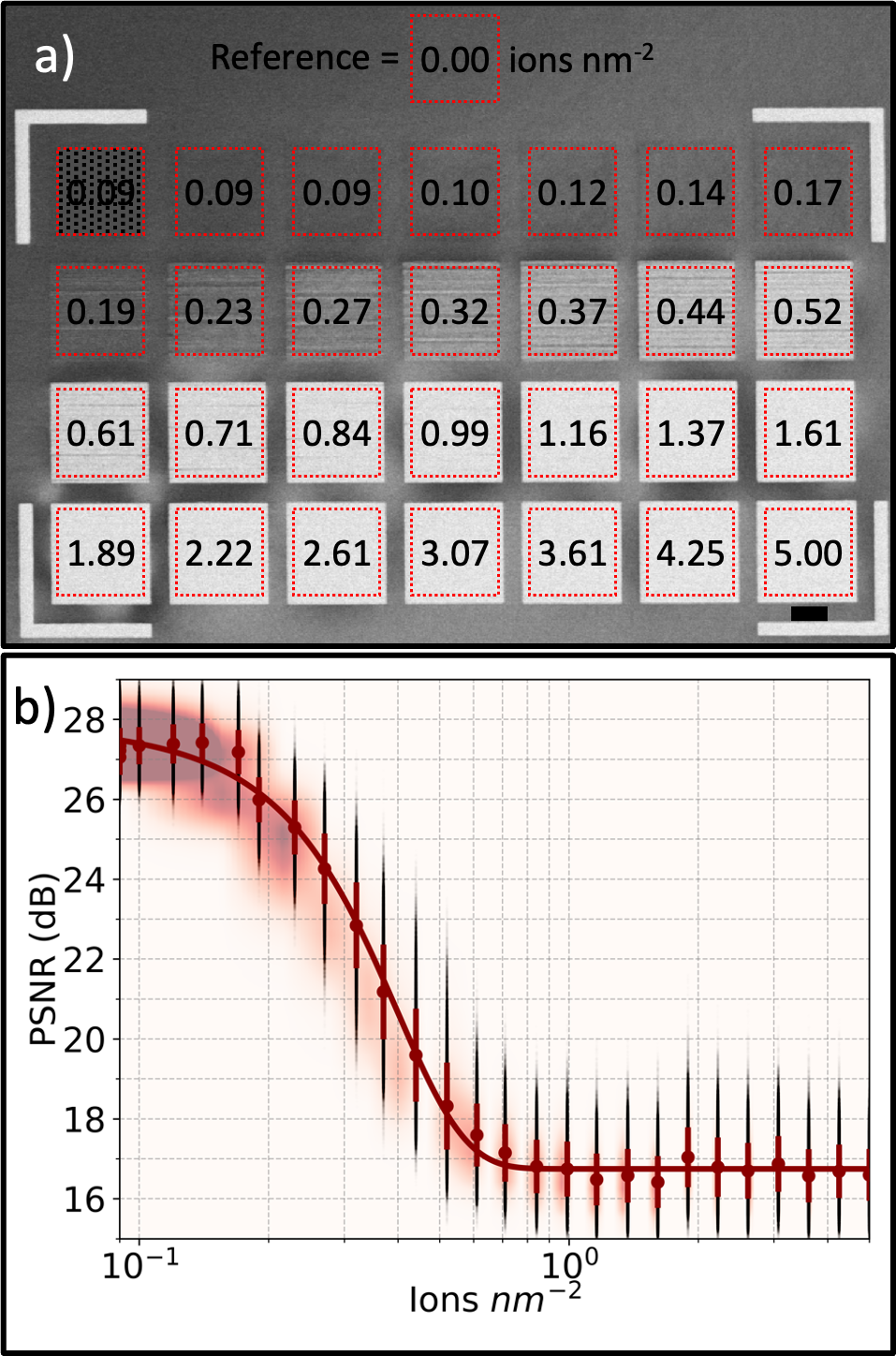}
    \caption{Calibration of peak signal-to-noise ratio (PSNR) as a function of ion dose. a) Low-angle Annular dark-field STEM (LAADF-STEM) image of the calibration region, where the numbers indicate the ion implantation dose in ions per nm². Red dashed squares mark the \SI{4}{\micro\meter\squared} areas scanned to acquire the 4D-STEM data used for PSNR analysis. b) PSNR values plotted as a function of ion dose. Black dots represent individual PSNR values (100,000 per dose) calculated from the CBED patterns in each calibration region. Red dots show the mean PSNR for each dose, with vertical bars indicating the standard deviation. Reddish blurred regions underneath the plot corresponds to heat distribution of the PSNR values for each dose. The red curve is a fit of the data using a Gaussian error function.}
    \label{fig:Calibration}
\end{figure}

\Cref{fig:CBED_series} displays a series of convergent beam electron diffraction (CBED) patterns acquired using 100 keV electron precession---a tilted electron beam rotation around the optical axis producing a quasi-kinematical diffraction \cite{Eggeman2012}---from the calibration regions subjected to varying ion implantation doses. To enhance visualisation, each pattern represents the integration of 10,000 diffraction patterns (DPs) collected from the corresponding calibration square shown in \cref{fig:Calibration}(a). The numerical values in the top-left corner of each pattern indicate the ion dose in ions cm$^{-2}$, ranging from 0.00 (pristine) to 5.00~ions cm$^{-2}$. As the dose increases, a progressive degradation in the CBED quality is observed, culminating in complete amorphisation. From those images, it is clear that CBED patterns offer a continuous greyscale intensity distribution across the entire diffraction pattern, in contrast to the binary-like contrast of spot patterns obtained at lower convergence semi-angles (parallel beam). This richer intensity variation enhances the sensitivity and robustness of peak signal-to-noise ratio (PSNR) analysis \cite{Qasim2018,Nayak2021} to enable a highly sensitive calibration of the ion dose based on the degradation in the CBED patterns. See supplementary materials for details of the PSNR methodology.

CBED patterns are highly sensitive to experimental conditions, alignments precision, and beam coherence; which combined with the logarithmic sensitivity of PSNR to small intensity differences, justify the lower observed values even in nearly undamaged regions. Importantly, the use of PSNR in this context is not to assess visual fidelity, as in image compression, but rather to quantify structural deviation in the reciprocal plane. In crystalline materials, CBED patterns emerge from the coherent elastic scattering of a convergent electron probe by the periodic atomic lattice, resulting in well-defined diffraction discs that exhibit characteristic symmetry and intricate fine structures. When ion implantation occurs, it introduces lattice disorder and partial amorphisation, disrupting long-range order and enhancing diffuse scattering, which in turn redistributes intensity across the pattern. Although CBED is not strictly governed by the Bragg condition, due to the angular spread of the incident beam, the diffraction contrast remains highly sensitive to the integrity of periodic structures \cite{Williams1996}. As a result, PSNR serves as a robust quantitative metric to detect deviations from the pristine diffraction signature. High PSNR values indicate strong similarity to the undamaged crystalline reference, suggesting minimal structural disruption, whereas lower values reflect increased deviation, consistent with greater lattice disorder.

\begin{figure*}[htbp]
    \centering
    \includegraphics[width=\linewidth]{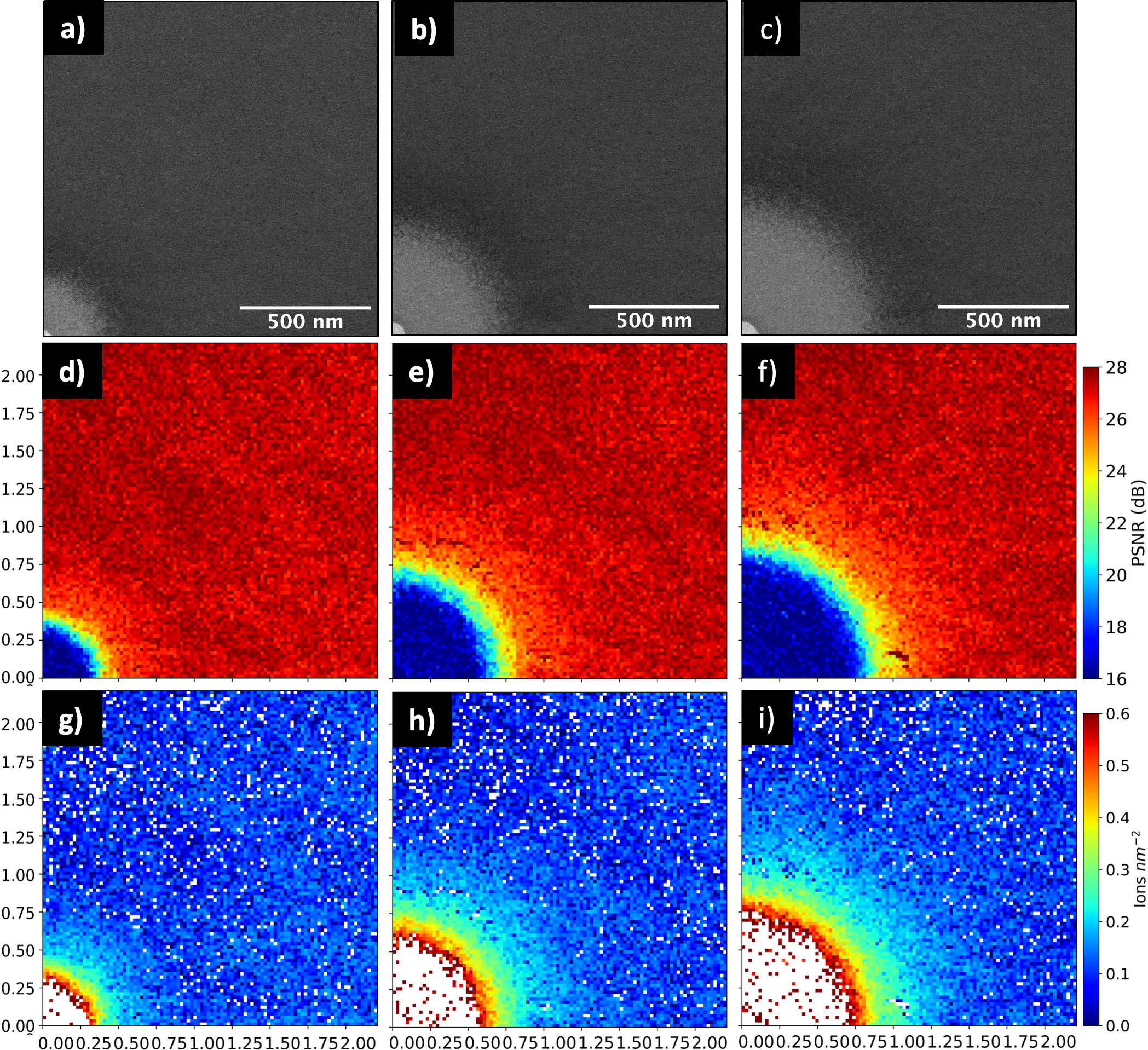}
    \caption{Focused ion beam-induced damage analysis quadrants with its bottom-left corner aligned at the centre of a milled hole. a–c) ADF-STEM images illustrating the damage across the quadrant with contrast defined by scattering events. d-f) Averaged PSNR values obtained by comparing each $100\times100$ scanned pixels from the 4D-STEM dataset to 10 randomly selected CBED patterns from a pristine reference region. g–i) 2D ion dosage maps derived by converting the PSNR values to ion dose (ions nm$^{-2}$) using parameters from the Gaussian error function fit. White regions indicate PSNR values outside the plateaus in \cref{fig:Calibration}, corresponding to either minimal or saturated damage levels. The left, middle, and right columns correspond to dwell times of 1~s, 5~s, and 10~s, respectively.}
    \label{fig:quadrants}
\end{figure*}

Based on the PSNR analyis of the CBED patterns obtained from the regions marked in \cref{fig:Calibration}(a), a plot of PSNR as a function of ion dosage could be obtained and is depicted in \cref{fig:Calibration}(b) with the abscissa plotted in a $\log_{10}$ scale. The PSNR dependence follows a smooth sigmoidal transition between two asymptotic states, being the regime at higher dosage consistent with the expected amorphisation of the silicon single-crystal and the lowest doses approaching a regime in which just a few more ions per \SI{}{\nano\meter\squared} do not contribute to additional amorphisation at the detection limit of the technique. To model the smooth transition observed in the dataset, we employed a sigmoidal fitting function based on the Gauss error function, defined as:
\begin{equation}
    f(x) = a \cdot \operatorname{erf}\left(b(x - c)\right) + d,
\end{equation}
where \( \operatorname{erf}(x)\) is the standard error function, and \( a \), \( b \), \( c \), and \( d \) are fitting parameters.

SRIM/TRIM \cite{Ziegler1985} and SUSPRE \cite{Webb1986} simulations suggest that amorphisation in silicon implanted with 30~keV Ga$^+$ ions occurs at a dose of approximately 5~ions per \SI{}{\nano\meter\squared}. However, our PSNR measurements indicate that the onset of amorphisation may occur at significantly lower doses, up to five times smaller than predicted. This discrepancy arises because these simulations typically define amorphisation as the point at which each ion collision displaces a host atom. In reality, crystallinity is not a binary property, and long-range order can be substantially disrupted well before complete atomic displacement occurs. A crystal cannot be considered ``partially amorphous'' in a straightforward way, hence a crystal cannot meaningfully be described as 50\% amorphous, as the long-range periodicity of the lattice is already disrupted at that point. A more nuanced understanding of this transition will be explored in a forthcoming study, combining molecular dynamics simulations \cite{Liu2019} with the experimental 4D-STEM measurements.

\begin{figure*}[t]
    \centering
    \includegraphics[width=0.8\linewidth]{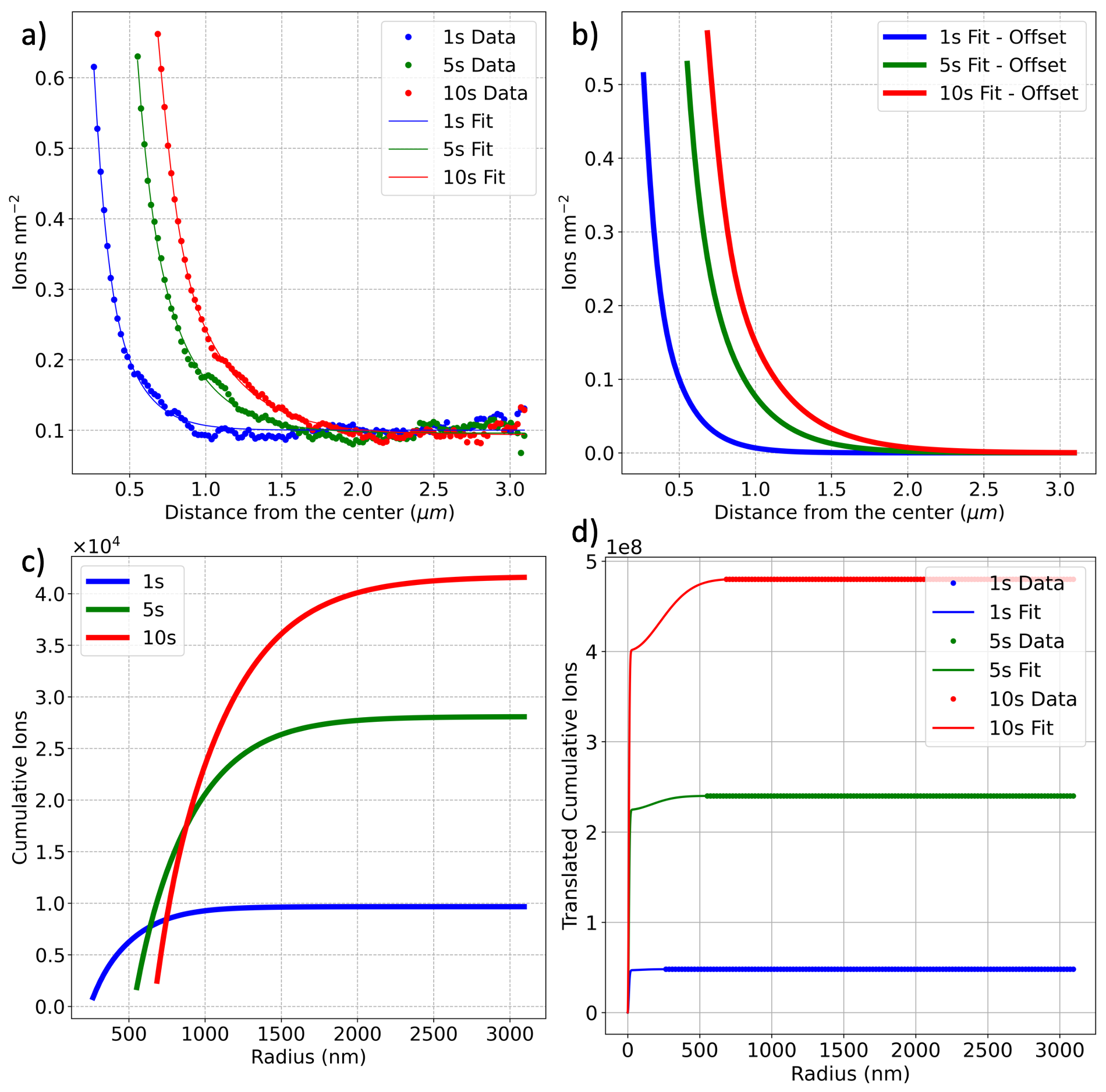}
    \caption{Procedure for obtaining the radial probability distribution of implantation events. a) The quantifiable ion density range from \cref{fig:Calibration} (in ions nm$^{-2}$) was fitted using a combination of a Gaussian and an exponential function. b) Fitted ion density profiles. c) Fitted curves integrated to obtain the cumulative number of implanted ions as a function of radial distance. The non-zero starting point reflects the exclusion of fully amorphous and hole regions, where the majority of ions are initially deposited. d) The far-right end of the cumulative curves was offset to match the total expected number of implanted ions based on a 7.5~pA beam dwelled over times of 1~s (blue), 5~s (green), and 10~s (red). To fill the unquantified gap, a combination of two Gaussian components and an exponential tail was employed: the inner Gaussian was constrained to a fixed FWHM of $\approx15$~nm, representing the imaging resolution core, while the middle Gaussian---representing secondary scattering---was allowed to vary in FWHM during fitting.}
    \label{fig:fitting}
\end{figure*}

\begin{figure}[htp!]
    \centering
    \includegraphics[width=\linewidth]{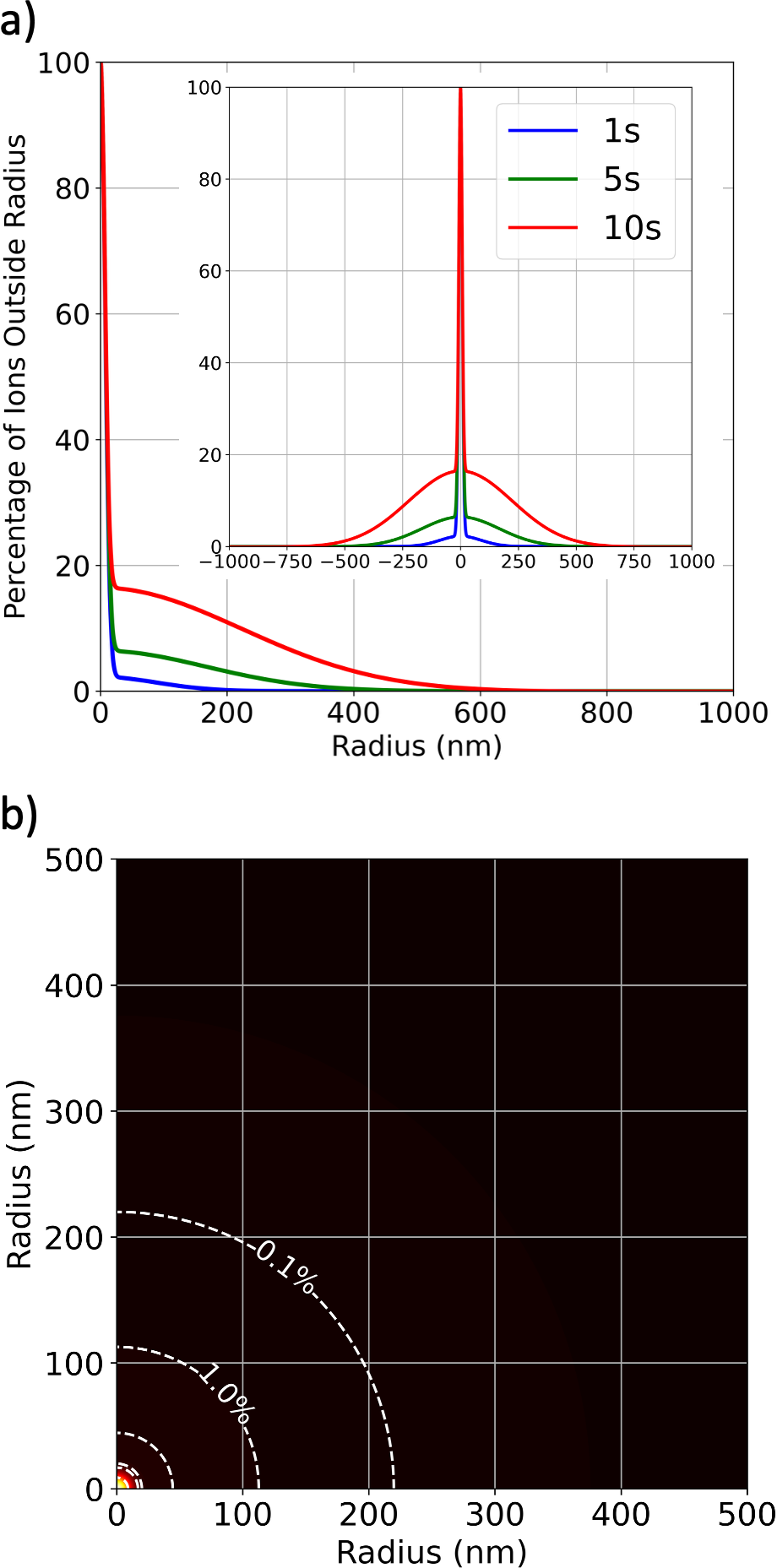}
    \caption{Radial probability contour plots derived from the integrated area under the fitted curves in \cref{fig:fitting}, combined with the known ion dose delivered during irradiation. (a) Displays a plot of percentage of ions outside a given radius. The inset is a vertically-mirrored plot to illustrate the expected damage profile around and across a FIB-dwelled spot. Panel (b) shows a two-dimensional heat map representing the probability of finding ions beyond a given radius for the 1~s dwell time spot (total incident ions equals to $4.68\times10^7$). The probability values are indicated by the dashed white contours over the plot.}   
    \label{fig:percentage}
\end{figure}

\Cref{fig:quadrants} displays images of a quadrant where an ion beam is held for (1~s, 5~s, and 10~s), with the  beam centre positioned at the bottom left in each panel. The upper three images (a-c) are low-angle annular dark-field (LAADF-STEM) images, the middle set (d-f) depicts two-dimensional (2D) PSNR maps, and the bottom trio (g-i) illustrates the calculated ion dosage surrounding the FIB-milled hole. As previously discussed, LAADF micrographs are highly sensitive to beam-induced damage, and it is evident that gallium ions reach distances of up to approximately 500~nm from the irradiation site. A quantitive measure of this damage is obtained from the PSNR images of the same field of view. Here, each CBED pattern was compared to ten randomly selected reference patterns from an undamaged region, as described in the supplementary materials, and the mean PSNR was calculated. Using the fit parameters derived from the calibration curve in \cref{fig:Calibration}b, a 2D heat map of ion density was reconstructed, illustrating the spatial distribution of defects around the beam incidence point. White pixels in the ion distribution maps indicate regions where the PSNR falls below the baseline value, corresponding to ion doses exceeding 5~ions nm$^{-2}$. These areas are either fully amorphised or represent locations where the 30~nm thick membrane has been perforated. Although the region away from the damage shows a small ion density (<0.1~ions nm$^{-2}$), this highlights the high sensitivity of 4D-STEM to subtle CBED pattern variations likely due to slight off-axis probing considering the relatively long distance to which reference patterns were collected (tens of microns). Importantly, the uniform signal across the area confirms the absence of scanning artefacts, unlike ADF imaging over large fields of view.

To visualise the full point spread function (PSF) of the FIB-induced damage, the PSNR maps in \cref{fig:quadrants}(d-f) were symmetrically mirrored in two steps: first vertically to the left, and then horizontally downward. This process effectively reconstructs a full radial profile centred on the beam impact point, representing a complete damage distribution with a damage profile described by \cref{Eq.BeamProfile} in the supplementary materials (SM). \Cref{fig:1_over_PSNR} in SM shows the profiles acquired in a low-dose regime, from 0.09~ions nm$^{-2}$ to 5~ions nm$^{-2}$. This contrasts significantly with the higher dose regimes used in the study by Drezner et al. \cite{Drezner2017}, where TEM experiments and TRIM modelling were employed to characterise amorphisation in single-crystal silicon at doses exceeding 10–100 ions nm$^{-2}$. Due to limitations of the 4D-STEM technique at higher doses---particularly in regions with hole formation---the PSNR$^{-1}$ plot exhibits certain flaws. A more detailed discussion of this representation is provided in the supplementary materials.

\Cref{fig:fitting}(a) plots the local ion dose (ions nm$^{-2}$) as a function of radial distance from the beam centre for three Ga$^+$ dwell times (1~s, 5~s, and 10~s). These profiles were extracted from regions where the ion dose remained below the amorphisation threshold, enabling reliable quantitative analysis. To capture the full radial ion distribution, \Cref{fig:fitting}(b) shows fitted models comprising a broad Gaussian combined with an exponential tail. \Cref{fig:fitting}(c) presents the cumulative ion dose as a function of radius, calculated from the fitted profiles. Because the saturated central region was excluded from the original dose measurements, these curves do not start at high values and underestimate the total ion dose near the beam centre. To address the missing contribution from the saturated core, \cref{fig:fitting}(d) shows the reconstructed cumulative ion curves obtained by fitting the radial dose profiles with a composite model consisting of two Gaussian functions and an exponential tail. The first Gaussian represents the tightly focused beam core and was fixed with a full width at half maximum (FWHM) of approximately 15~nm across all dwell times, consistent with the expected “imaging resolution”. The second Gaussian captures the broader mid-range distribution and was allowed to vary in width to account for dose-dependent broadening. The exponential component models the long-range tail and defect-mediated diffusion. The total integrated ion dose from the fits was constrained to match the known number of ions delivered for each dwell time, based on the beam current (7.5~pA) and exposure durations. This approach ensures that the reconstructed cumulative curves begin as zero at the beam centre and accurately capture the full spatial distribution of ions, including contributions that were not measurable directly due to detector saturation.

\Cref{fig:percentage} translates the reconstructed dose profiles into radial probability distributions, showing the percentage of ions expected to land beyond a given radius. The mirrored plot in the inset of \cref{fig:percentage}a clearly shows that a FIB damage profile is a convolution of three different components as expected from theoretical calculations \cite{Drezner2017}. The radial probability plots confirm that the core remains spatially confined, while the scattering tail increasingly dominates the long-range distribution at extended distances. \Cref{fig:percentage}b presents a two-dimensional heat map illustrating the probability distribution of ion damage events as a function of radial distance for the spot dwelled for 1~s. This distinction has important implications for different FIB applications. In deterministic single-ion implantation, the goal is to place exactly one ion per site with nanometre-scale precision. Extrapolating from the measured distributions, the radial spread for a single-ion event is expected to be extremely narrow---on the order of a few nanometres---due to the absence of cumulative scattering effects. This makes crosstalk negligible in arrays with pitches as small as 20–50~nm. In contrast, when using the focused ion beam for imaging fiducial marks prior to implantation, the dwell times often fall within the 1~s to 10~s range studied here. Under these conditions, the spread of ions becomes more significant, and the scattering tails can extend far beyond the core region. In these scenarios, precision and careful planning becomes critical to avoid unintentional damage to nearby structures. This highlights the importance of being able to map the point spread function of the beam and tailoring beam parameters to the spatial constraints of the device. In contrast, for higher dose applications such as lamella preparation and circuit repairs, the lateral extent of the amorphisation layer is a critical parameter. The exponential tail in the ion distribution contributes to damage well beyond the beam centre and means to mitigate the lateral damage footprint must be incorporated in the process.

The supplementary material demonstrates the robustness of combining 4D-STEM with PSNR analysis by significantly varying beam and camera acquisition parameters, while still achieving ultra-sensitive defect detection and potential spatial resolution of few angstroms. Additionally, it explores the potential of electron backscatter diffraction (EBSD) as an alternative approach for achieving comparable fidelity in focused ion beam (FIB) profiling.

\section*{Conclusions}
The primary objective of establishing the relationship between PSNR and ion dose is to enable reproducible, quantitative mapping of damage around localised ion beam exposures---particularly in regions where conventional imaging techniques fall short. We demonstrate how to convert PSNR values obtained from regions surrounding sites where the FIB was held stationary into spatially-resolved ion dose estimates. PSNR metrics require a robust dataset with spread-out pixels intensity that are intrinsically related to the crystal periodicity; which is met by using 4D-STEM hypercubic datasets of CBED patterns. The approach allows to reconstruct the two-dimensional distribution of implanted ions across and around the core impact region. Notably, the peripheral zone which is often overlooked in ion beam damage studies due to the lack of techniques that combine high sensitivity with nanoscale spatial resolution. Our method addresses a critical gap in knowledge by offering a new pathway for characterising subtle damage gradients in ion-irradiated materials, thereby enabling direct comparisons with implantation models and establishing a foundation for correlating structural damage with functional degradation in nanoscale devices.

\section*{Supplementary Material}

\setcounter{figure}{0}
\renewcommand{\thesection}{SM\arabic{section}}
\renewcommand{\thefigure}{S\arabic{figure}}
\renewcommand{\theequation}{S\arabic{equation}}
\setcounter{equation}{0}
\bibliographystyle{unsrt}


\section{Correction of Optical Off-Axis Artefacts in ADF-STEM Imaging} \label{sec:AF-STEM}
\begin{figure}[b]
    \centering
    \includegraphics[width=\linewidth]{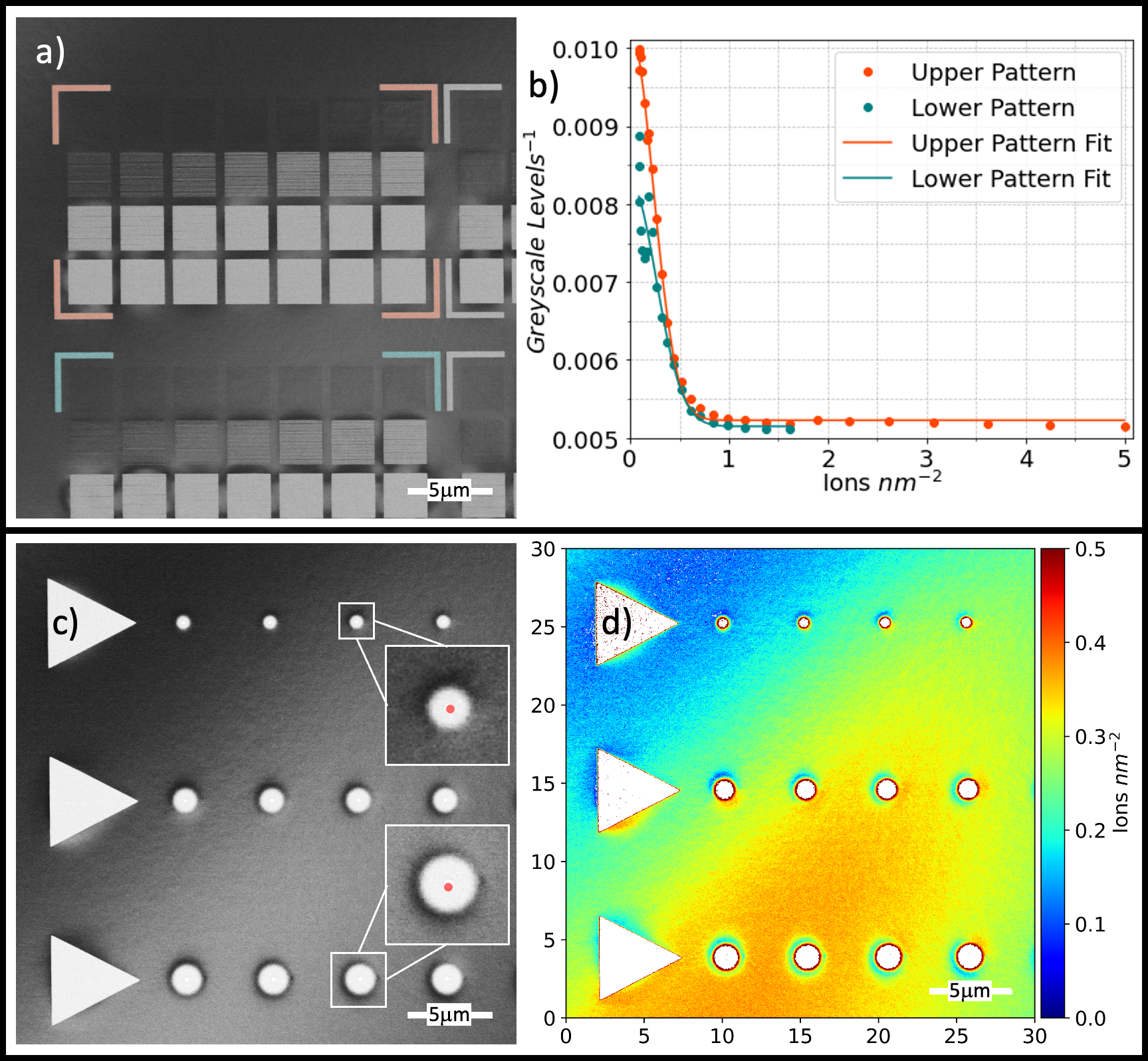}
    \caption{Tentative of damage quantification using LAADF-STEM signal. a) Calibration $2.5 \times 2.5$\SI{}{\micro\meter\squared} squares created by 30~keV Ga$^+$ implantation with doses exponentially ranging from 0.09~ions nm$^-2$ (repeated three times) to 5~ions nm$^{-2}$ (from darker to lighter tones as explained previously), contained within the fiducial marks highlighted in either orange or green. b) One-over-greyscale levels plotted as a function of the calibration doses. Levels were obtained by averaging the intensity of individual pixels over a 4 $\mu$$m^2$ area within each square. c) LAADF-STEM image obtained at 100 keV for three sets of dwelled holes (1~ms, 5~ms, and 10~ms from top to bottom. Insets highlight the holes’ area in red, surrounded by damaged silicon. d) 2D heat map of expected dose as a function of spatial position obtained by cross correlating the pixels’ bit intensities in (c) to the orange calibration curve in (b). Regions marked in white correspond to doses exceeding 5~ions nm$^{-2}$.}
    \label{fig:damage_calib}
\end{figure}

We designed an experiment in which twenty-eight $2.5 \times 2.5$\SI{}{\micro\meter\squared} squares were implanted with a 30~keV Ga$^+$ beam at doses ranging from 0.09~ions nm$^{-2}$ to 5~ions nm$^{-2}$ (\cref{fig:damage_calib}a). By imaging these squares using low-angle annular dark-field (LAADF-STEM), we can produce a calibration curve of greyscale levels as a function of ion dose (\cref{fig:damage_calib}b), which will be correlated with a scan around ion-milled holes. It is important to note that when imaging in STEM mode using LAADF detection, the events are detected just outside the bright-field cone, collecting electrons that have been coherently scattered at low-angles (e.g. slightly damaged regions) or diffuse scattering from the near-forward electrons through the amorphous regions (i.e., reduced channelling) \cite{Nellist2010}.

In \cref{fig:damage_calib}c, the row of holes on the right of the triangles (fiducial marks) were dwelled at 1~ms, 5~ms, and 10~ms from top to bottom, respectively, with the insets’ region highlighted in red indicating where a hole is located. During LAADF-STEM imaging of the implanted squares, we observed a systematic greyscale offset across the scan area. This artefact arises from optical off-axis effects, which persist even when operating in descan mode with a two-stage deflection coil. The artefact manifests as a gradient in image intensity, causing squares with identical ion doses to appear with different greyscale levels, as highlighted by the offset between the orange and green curves in \cref{fig:damage_calib}b. The origin of this effect is attributed to imperfect beam steering and minor misalignments of the optical axis relative to the crystal zone axis and, finally, to the detector annulus. At large collection angles, ADF contrast is dominated by Rutherford-like scattering proportional to $Z^{1.7-2}$; whereas at low angles diffraction, the contrast transfer function oscillations with defocus and aberrations become increasingly important. Any residual optical offset leads to spatial variations in scattering angle and collection efficiency, thus producing a continuous intensity gradient. To mitigate these artefacts, all calibration squares and regions of interest must be acquired within a single field of view--- and as small as possible---. Additionally, post-acquisition processing such as background subtraction or normalisation using reference regions can reduce the effect of the offset gradient. However, these methods cannot fully account for local variations in contrast transfer. For this reason, we proposed a 4D-STEM experiment where a full diffraction pattern is recorded at every probe position. Such a dataset would enable centre-of-mass correction and virtual detectors post-processing, providing a more robust and quantitative measurement of ion-induced damage.

\section{PSNR Analysis}\label{sec:PSNRAnalysis}

\begin{figure}
    \centering
    \includegraphics[width=\linewidth]{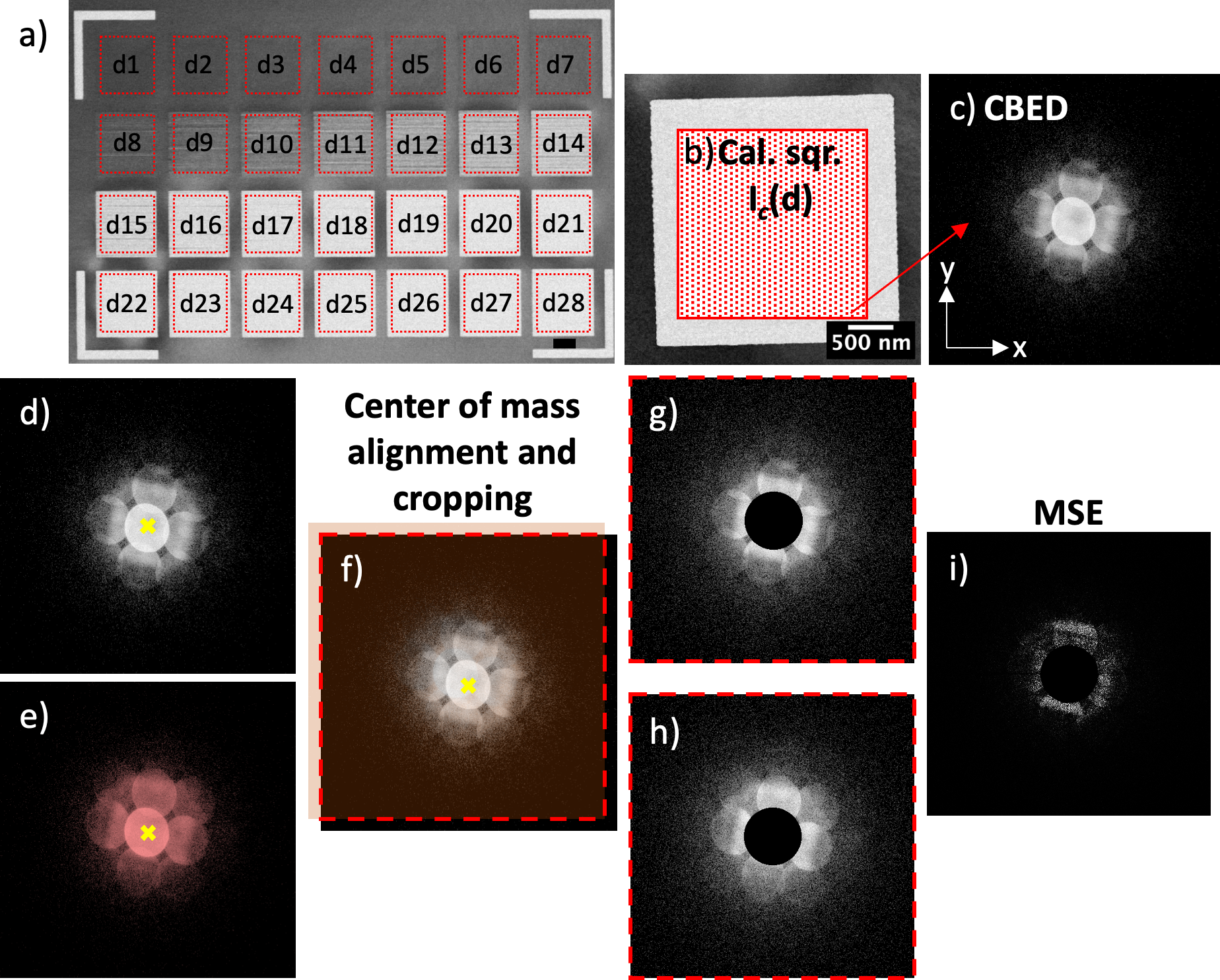}
    \caption{Workflow for quantifying damage in single-crystal silicon using convergent beam electron diffraction (CBED) and peak signal-to-noise ratio (PSNR) metrics. (a) Scanning electron microscopy (SEM) secondary electron image of a 30~nm-thick single-crystal Si membrane implanted with 30~keV Ga$^+$ ions. Dashed-line rectangles correspond to scanned regions inside areas implanted with a known ion dosage increasing from d1= 0.09~ions nm$^{-2}$ to d28= 5.00~ions nm$^{-2}$. b) High-magnification annular dark-field STEM (ADF-STEM) image of one calibration square, illustrating the $100\times100$ CBED patterns (DPs) acquired over a \SI{4}{\micro\meter\squared} area. c) Example CBED pattern from this dataset. d, e) Representative DPs from a reference (unimplanted) region just above the squares in (a) and from one of the implanted regions, respectively. Yellow crosses mark the centre-of-mass in each DP. f) Overlay of the reference and implanted DPs after aligning their centres-of-mass. Pixels outside the red dashed square are cropped to standardise the array shape, generating conformable arrays like (g) and (h), prior to mean squared error (MSE, in (i)) and PSNR calculations.}
    \label{fig:CBED_schematic}
\end{figure}

Using dynamic precession electron diffraction, a quasi-kinematical dataset comprising a broad range of measured reflections was acquired at each dwell point. The implanted squares highlighted by red dashed squares in \cref{fig:CBED_schematic}a served as calibration regions (similarly to \cref{fig:damage_calib}a), \(I_c(d)\). A total of $n=10,000$ pixels were scanned over a \SI{2}{\micro\meter} $\times$ \SI{2}{\micro\meter} area, deliberately avoiding the edges (\cref{fig:CBED_schematic}b). Each dwell point produced a $512\times512$ pixel convergent beam electron diffraction (CBED) pattern (\cref{fig:CBED_schematic}c). The analysis was based on the peak signal-to-noise ratio (PSNR). For this, each CBED pattern from the implanted regions was compared to 10 randomly selected CBED patterns (from the 10,000 acquired) in a pristine area adjacent to the calibration squares region in \cref{fig:CBED_schematic}a. In total, 2,800,000 comparisons were performed---100,000 for each of the 28 implanted squares with ion dose $d$ ions per nm.

The pixel-wise standard deviation across all 10,000 reference patterns was calculated as:
\begin{equation}
    \sigma_r(x, y) = \sqrt{\frac{1}{n} \sum_{j=1}^{n}\left( I_r^{(j)}(x, y) - \overline{I_r}(x, y) \right)^2},
\end{equation}
where \( I_r^{(j)}(x, y) \) is the intensity at pixel \((x, y)\) in the $j^\text{th}$ diffraction pattern, and \( \overline{I_r}(x, y) \) is the mean intensity at that pixel across all \( n = 10,\!000 \) reference patterns. The maximum observed standard deviation was approximately 8 intensity units. Given the 16-bit grayscale range of the CBED patterns \((\mathrm{MAX}_I = 2^{16} - 1 = 65,\!535)\), this corresponds to a relative variation of less than 0.013\% of the expected full intensity range, indicating low variability. Therefore, a subset of 10 reference patterns was deemed sufficient for PSNR calculations without compromising accuracy while maintaining realistic processing times.

For each comparison between reference and target patterns, the centre of mass was calculated, \cref{fig:CBED_schematic}(d-e), and aligned via offsetting. Excess edges were trimmed, outside of dashed lines in \cref{fig:CBED_schematic}f, to ensure alignment, and the resulting dimensions of each trimmed target/reference pair were recorded. At the end of the loop, all pairs were resized to match the smallest dimensions, $M\times N$, by trimming random edges, \cref{fig:CBED_schematic}(g-h), ensuring uniformity across the dataset.

The mean squared error (MSE) between each calibration pattern \( I_c^{(i)}(x, y) \) and 10 random pixels from the reference region $I_{r,\text{rand}}(x, y)$ was calculated (\cref{fig:CBED_schematic}i) as
\begin{equation}
    \mathrm{MSE}(d)^{(i)} = \frac{1}{M N} \sum_{x=1}^{M} \sum_{y=1}^{N} \left( I_c(d)^{(i)}(x, y) - I_{r,\text{rand}}(x, y) \right)^2.
\end{equation}
The peak signal-to-noise ratio (PSNR) was then calculated as:
\begin{equation}
    \mathrm{PSNR}(d)^{(i)} = 10 \cdot \log_{10} \left( \frac{\left(\mathrm{MAX}_I(d)^{(i)}\right)^2}{\mathrm{MSE}(d)^{(i)}} \right),
\end{equation}
for each of the 10 MSE values for each pixel and dose, where \( \mathrm{MAX}_I(d)^{(i)} \) is the maximum bit intensity measured in \( I_c^{(i)}(x, y) \) for the $i^\text{th}$ pixel in the calibration square \(I_c(d)\) from \(i=1\) to \(i=n=100,000\). PSNR is expressed in decibels (dB), with higher values indicating greater similarity (i.e., lower noise or distortion). In this study, PSNR values across the calibration dataset ranged from approximately 15\,dB to 30\,dB. While PSNR values for 16-bit natural images typically range from 60 to 80\,dB in image compression literature~\cite{Welstead1999}, the values observed here reflect the fundamentally different nature of the data.

\section{Preliminary Beam Profile Estimation via PSNR$^{-1}$ Radial Mapping}\label{sec:1-over-PSNR}

\begin{figure}[t]
    \centering
    \includegraphics[width=\linewidth]{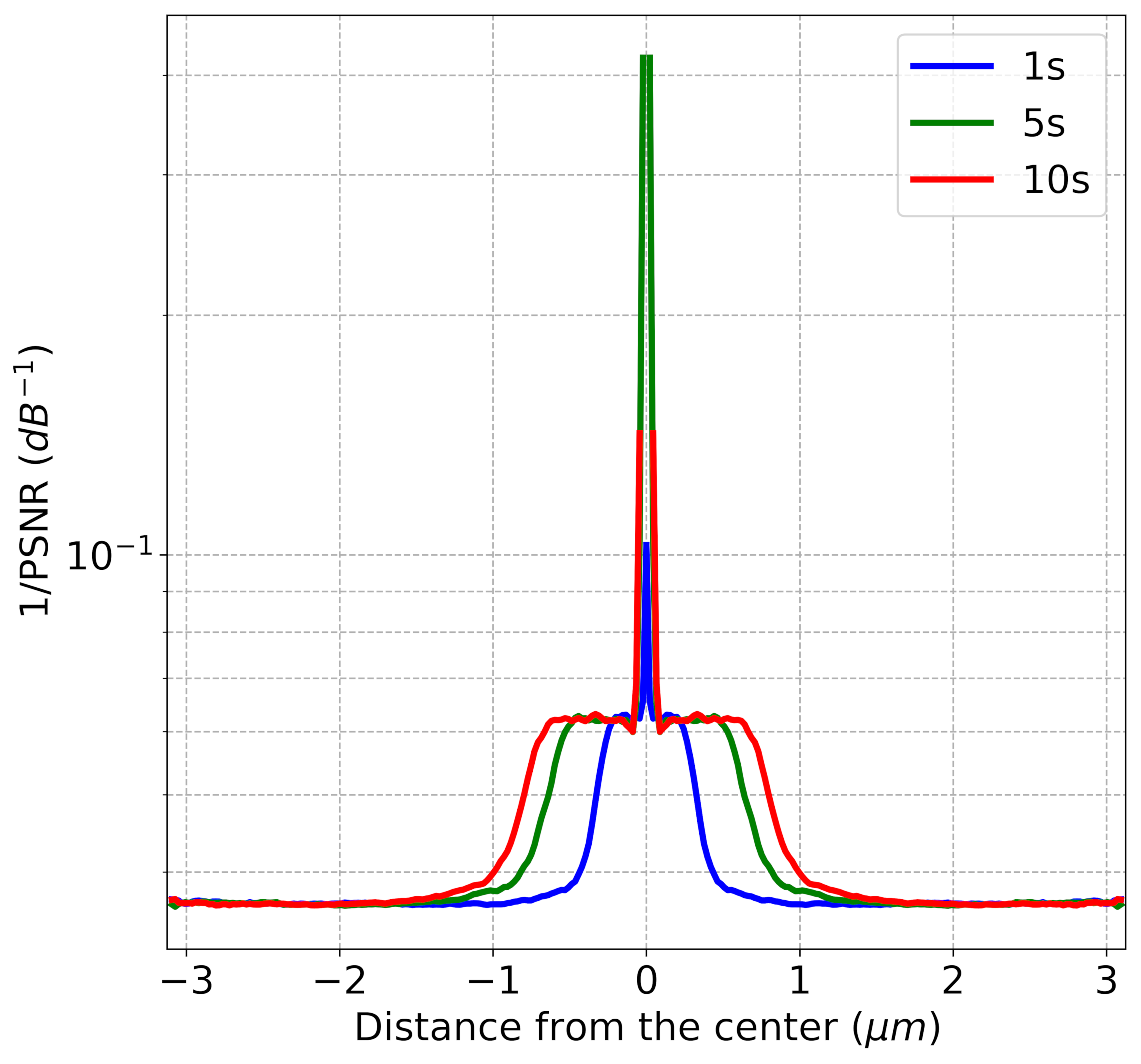}
    \caption{Point spread function (PSF) preliminary profiles obtained by applying two successive mirroring operations to reconstruct a full circle from a single quadrant (shown in \cref{fig:quadrants}d-f), enabling the extraction of radial intensity profiles represented in terms of 1 over PSNR in logarithmic scale. The resulting curves were interpreted as a combination of two Gaussian functions and an exponential tail, with the Gaussians saturation attributed to the limitations of the PSNR method in quantifying degrees of amorphousness after crystal symmetry is lost.}
    \label{fig:1_over_PSNR}
\end{figure}

The shape of the PSNR$^{-1}$ curves (\cref{fig:1_over_PSNR} in our data resembles the damage profiles reported in Figure 6 of Drezner et al. \cite{Drezner2017}, which were modelled using a combination of two Gaussian functions and an exponential decay:
\begin{equation}
    D(r) = A_1 e^{-\frac{r^2}{2\sigma_1^2}} + A_2 e^{-\frac{r^2}{2\sigma_2^2}} + A_3 e^{-\lambda r} +C,
    \label{Eq.BeamProfile}
\end{equation}
where D(r) represents the damage intensity---represented here as the inverse of the PSNR, based on the established correlation that lower PSNR values correspond to higher degrees of amorphisation---as a function of radial distance (r), with  $A_1$, $A_2$, and $A_3$ as amplitude coefficients, $\sigma$$_1$ and $\sigma$$_2$, as the standard deviations of the two Gaussian components, $\lambda$ as the decay constant of the exponential tail, and $C$ as a constant to correct for the non-zero ion density at regions far from the implantation spot. In this composite model, the narrow Gaussian reflects the high-dose core focused along the FIB optical axis, effectively corresponding to the imaging resolution. The broader Gaussian accounts for non-focused neutral ions, intermediate-range scattering and the spatial spread of ion-induced cascades---including subsurface straggling, low-angle recoil trajectories and lateral channelling effects---; while the exponential term should account for long-range scattering at low-vacuum regimes or radiation-induced diffusion processes.

\begin{figure}[t]
    \centering
    \includegraphics[width=\linewidth]{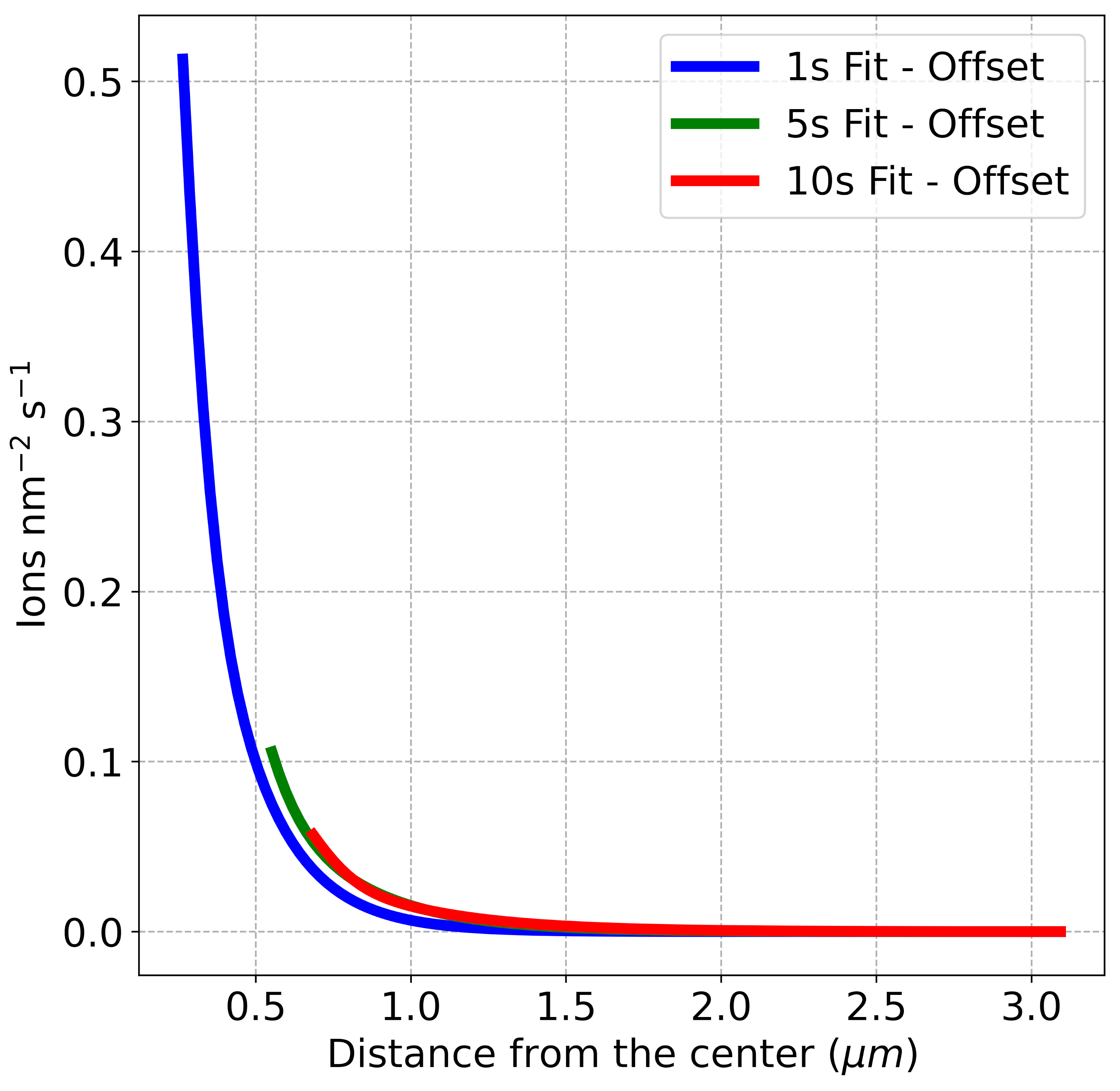}
    \caption{Time-normalised ion density profiles. Each curve was obtained by fitting a broad Gaussian combined with an exponential decay to the quantifiable regime (prior to amorphisation or through-thickness milling) around the FIB-dwelled spots, followed by normalisation to the respective dwell time.}
    \label{fig:saturation}
\end{figure}

Despite the qualitative agreement with the Drezner et al. \cite{Drezner2017} model, we observe deviations in the peak intensity and relative width of the fitted components. Notably, the amplitude of the narrow Gaussian appears reduced, and the broader Gaussian flattens toward the centre. This discrepancy occurs from PSNR saturation in regions where the local dose exceeds ~0.8 ions nm$^{-2}$, beyond which Bragg contrast is fully lost, and additional damage becomes indistinguishable via 4D-STEM. As a result, damage is systematically underestimated in the high-dose core. Importantly, we also find that the full width at half maximum (FWHM) of the damage profiles does not scale linearly with dwell time. For example, the FWHM of the 1~s profile is not one-fifth or one-tenth the size of those corresponding to 5~s or 10~s exposures. This non-linear scaling highlights fundamental characteristics of focused ion beam interactions: while the number of delivered ions increases linearly with dwell time, the lateral extent of damage broadens sub linearly due to several factors. First, ion–solid interactions are probabilistic and scattering events are governed by fixed angular distributions, meaning that the lateral range of displaced atoms is bounded. Second, at higher doses, collision cascades begin to overlap, and newly displaced atoms may recombine with existing vacancies or become trapped in pre-damaged regions, limiting further lateral propagation. Third, amorphisation itself modifies the material structure, potentially reducing channelling efficiency and changing ion penetration and scattering dynamics. Finally, cumulative thermal effects and radiation-enhanced diffusion may contribute to broader profiles at longer dwell times, but only weakly and non-linearly. Altogether, these mechanisms should lead to a saturation in damage width, consistent with prior findings in TRIM \cite{Ziegler1985} simulations, and better illustrated in \cref{fig:saturation}. The plots in \cref{fig:saturation} were obtained by normalising the fits from \cref{fig:fitting}b by the respective dwell times (i.e., dividing the ion density by 1, 5, and 10 for increasing dwell times). Although uncertainties in amplitude may still influence the probability plots in \cref{fig:percentage}, the closer proximity of tail widths for the 5~s and 10~s curves suggest that the long-range damage distribution approaches saturation with increasing dwell times. These assumptions can be further supported through complementary modelling (e.g., Monte Carlo \cite{Morvan1997, Hossinger1999} or TRIDYN \cite{Mller1984, Mller1985}).

\section{Real–Reciprocal Space Trade-off in High Convergence Angle Experiments}\label{sec:Heisenberg}

\begin{figure}[b]
    \centering
    \includegraphics[width=\linewidth]{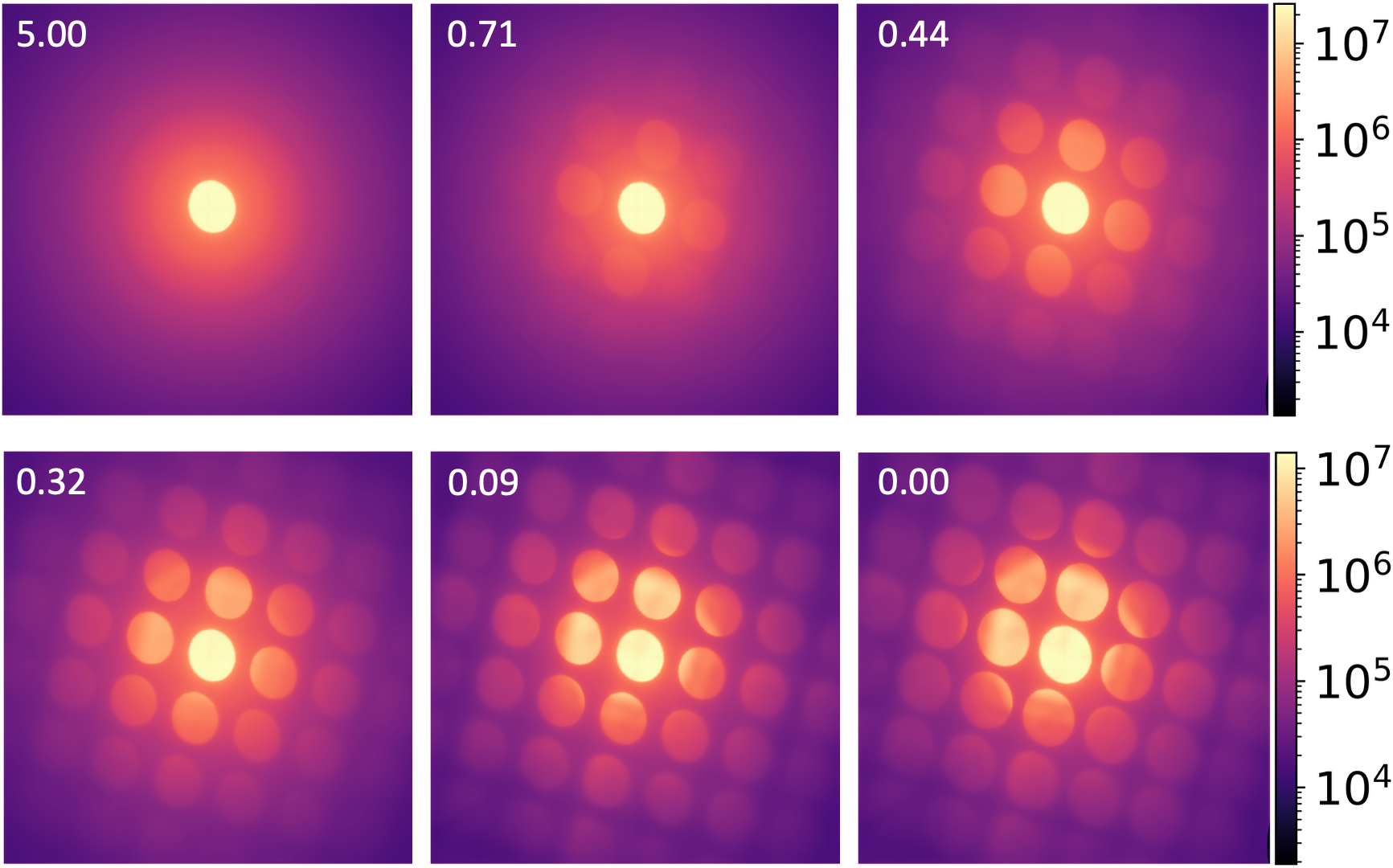}
    \caption{10,000 integrated CBED patterns obtained by scanning the 0.7~mrad precessed 100~keV, 8~mrad convergence semi-angle, 210~pA probe current with a 10~ms dwell time per pixel. The CBED patterns have dimensions of $512\times512$ pixels and are plotted using a logarithmic intensity scale. The numbers at the top-left of each CBED pattern indicate the dose used in the corresponding calibration square, with 0.00 representing the pristine region.}
    \label{fig:CBEDv2}
\end{figure}

To investigate the impact of probe parameters on 4D-STEM/PSNR output while mapping FIB PSFs, the convergence angle was reduced from 12~mrad to 8~mrad. A smaller convergence angle produces a larger real-space probe that could be a potential limitation, but in most 4D-STEM applications the scanning pitch size is larger than the probe diameter. Simultaneously, the reduced angular spread in reciprocal space results in sharper and more spatially confined CBED discs. This narrowing of the diffraction discs enhances the precision of centre-of-mass measurements, which in turn should improve the technique sensitivity to subtle lattice distortions and strain fields associated with defects. Moreover, the reduced CBED size minimises overlap between the central beam and higher-order diffraction features, lowering background interference and improving the interpretability of local intensity variations. These factors should collectively improve the accuracy and sensitivity of defect quantification in materials. Representative integrated CBED patterns acquired at lower convergence angles at same regions of those in \cref{fig:Calibration}) are presented in \cref{fig:CBEDv2}.

\begin{figure*}[htbp]
    \centering
    \includegraphics[width=0.8\linewidth]{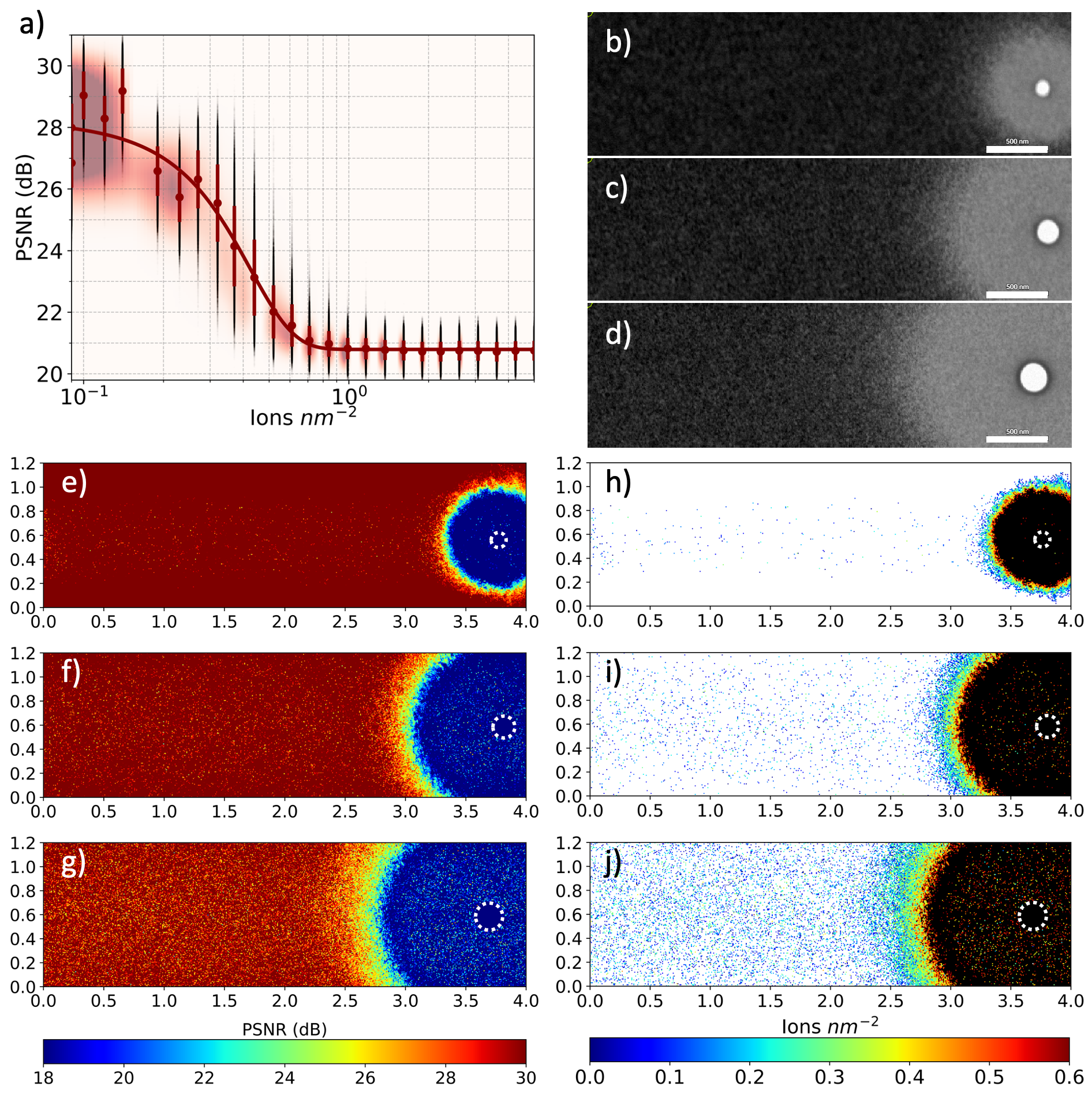}
    \caption{Overview of 4D-STEM and PSNR analysis for mapping the point spread function in focused ion beam-damaged regions. a) PSNR as a function of calibration ion dose for a new measurement set acquired with a smaller probe convergence semi-angle. Each black point represents one of 100,000 values collected per dose; red points indicate the median PSNR, with vertical bars denoting the error range. The red curve represents a Gaussian error function fit. b–d) ADF-STEM images of regions adjacent to holes milled using a 30 keV Ga$^+$ beam at 7.5~pA for 1~s, 5~s, and 10~s, respectively. e–g) 2D PSNR maps corresponding to the regions in (b–d), where each PSNR value was computed by comparing an individual CBED pattern to 10 randomly selected CBED patterns from a reference region, defined here as the far-left area just outside the field of views shown in b-d. Mean PSNR values were plotted. h–j) 2D ion density maps obtained by converting PSNR values to ion dose (in ions nm$^{-2}$) using parameters from the Gaussian error function fit. Regions marked with black dots represent areas below the lower PSNR plateau (i.e., heavily damaged), while white areas correspond to PSNR values indicative of doses below 0.09 ions nm$^{-2}$ (i.e., pristine regions). Dashed white circles indicate the locations of the milled holes.}
    \label{fig:Overview}
\end{figure*}

\Cref{fig:Overview} demonstrates the application of 4D-STEM with a higher convergence angle micro-probe to quantitatively map ion implantation dose around FIB-milled holes. In \cref{fig:Overview}(a), a calibration curve is established by comparing PSNR values from CBED patterns acquired in regions of known ion implantation dose to a reference region, as described previously. \Cref{fig:Overview}(b-d) show ADF images of regions surrounding holes milled with Ga$^+$ beam dwell times of 1~s, 5~s, and 10~s, respectively. These serve as the spatial context for the subsequent PSNR and ion density mapping. For each of these regions, a 4D-STEM scan was performed, and the CBED pattern at each pixel was compared to reference CBEDs obtained from the far-left periphery of the scanned area, assumed to be undamaged. The resulting PSNR maps are shown in \cref{fig:Overview}(e-g), where lower PSNR values (blue) indicate higher degrees of amorphisation. To translate these PSNR values into ion dose, the PSNR of each pixel was correlated to the calibration curve in \cref{fig:Overview}(a). The resulting 2D ion density maps are shown in \cref{fig:Overview}h–j, corresponding to the 1~s, 5~s, and 10~s dwell times, respectively. Note that pixels marked in black correspond to the vacuum region (hole) or the highly damaged (amorphous) regions. By selecting a reference region closer to the area of interest---leveraging the beam point spread function (PSF) characterised in the previous experiment (see \ref{fig:percentage})---we were able to correctly assign the region outside the beam core as defect-free without the need to add a constant to the fit of \cref{Eq.BeamProfile}.

Further investigation is needed to determine the limit of detection across different convergence semi-angles. One possible approach would be to use implanted vertical lines with reducing ion concentration, combined with a scan perpendicular to the lines---as in the methodology reported in \cite{Masteghin2024} for synchrotron X-ray nanodiffraction. Notably, this experiment demonstrates the robustness of the method by testing a distinctly different probe configuration, varying the position of the reference region, and adjusting analysis parameters. It also accounts for a slight zone-axis misalignment consistent across all scans. These variations did not significantly affect the resulting ion density maps, in contrast to conventional ADF imaging, which is more prone to artefacts.

\section{EBSD as a Potential Tool: Opportunities and Limitations to Be Explored}\label{sec:EBSD}

\begin{figure}[htbp]
    \centering
    \includegraphics[width=\linewidth]{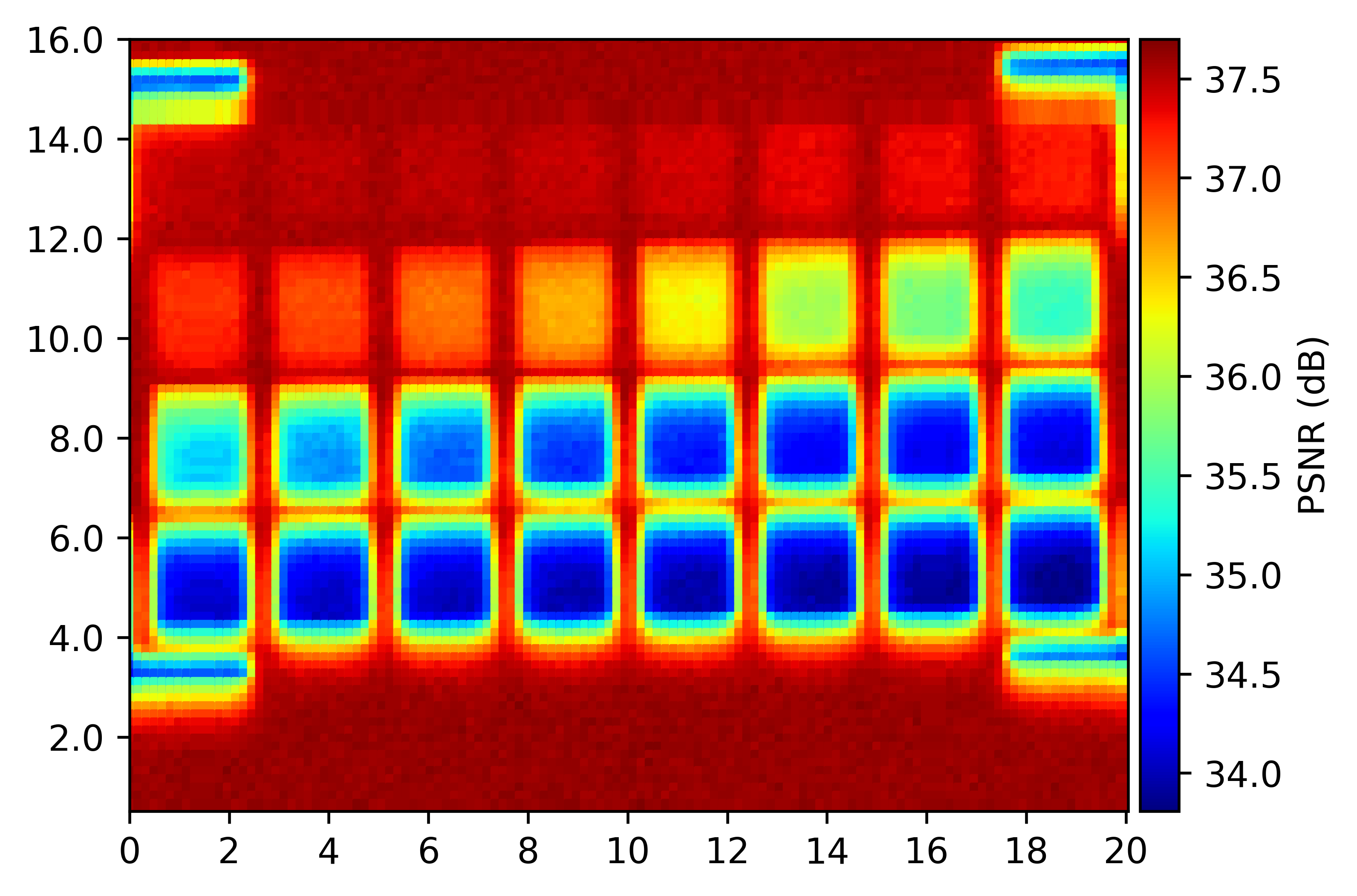}
    \caption{Electron backscatter diffraction (EBSD) data acquired from a region similar to that shown in \cref{fig:Calibration}(a). For simplicity, the PSNR analysis was performed without applying centre-of-mass alignment. Kikuchi patterns were compared to an averaged reference pattern obtained from a region located at the bottom of the imaged area. The diffuse edges observed in the maps are attributed to charging effects.}
    \label{fig:ESBD_V2}
\end{figure}

Finally, we note that electron backscatter diffraction (EBSD) can also be employed to investigate the point spread function of focused ion beams. EBSD can be performed directly on bulk samples, eliminating the need for electron-transparent lamella or membrane preparation, which is required for 4D-STEM. Additionally, the use of high beam currents in EBSD enables rapid data acquisition, provided that the sample can tolerate elevated electron doses and defects mobility under the electron beam is not a concern. Nevertheless, EBSD is inherently limited in spatial resolution compared to 4D-STEM. Its resolution, governed by the backscattered electron interaction volume (typically ~20~nm in Si at 20~keV), is insufficient to resolve closely spaced or singular defects, even if the overall sensitivity to local lattice distortions is comparable---a question still under investigation. By contrast, 4D-STEM, with its sub-nanoscale probe and precision diffraction-based measurements, offers superior spatial resolution and localisation capabilities, making it more suitable for quantifying fine-scale damage features. \Cref{fig:ESBD_V2} presents a representative PSNR map from EBSD, acquired in a region comparable to that shown in \cref{fig:Calibration}(a). The diffuse edges observed around the implantation sites are attributed to electrostatic charging effects, likely due to suboptimal grounding of the EBSD detector. Future studies will aim to systematically assess the detection limits of EBSD for defect quantification, particularly following a planned upgrade of the current system.

\clearpage

\section*{Acknowledgments}
MGM and SKC acknowledge financial support from the Engineering and Physical Sciences Research Council (EPSRC) (Grant No. EP/X018989/1). UoS-based authors acknowledge financial support from EPSRC (Grant No. EP/V036327/1, EP/W027070/1, and EP/X015491/1). The authors thank Silson Ltd for their support.

Portions of this manuscript, including textual refinement and code debugging, were assisted by GPT-5, a large language model developed by OpenAI.


\bibliography{references}


\end{document}